\newif\ifsingle

\ifsingle
\documentclass[11pt,draftclsnofoot, onecolumn]{IEEEtran}		
\else		
\documentclass[10pt,final, twocolumn]{IEEEtran}
\fi

\usepackage{Latent}

\SetKwInput{KwInput}{\textbf{Init}} 
\newcommand{\DiagSet}{\mySet{G}}

\IEEEoverridecommandlockouts 
 
\title{GSP-KalmanNet: Tracking Graph Signals via Neural-Aided Kalman Filtering}

\author{  
	\IEEEauthorblockN{Itay Buchnik, Guy Sagi, Nimrod Leinwand,  Yuval Loya, Nir Shlezinger,~\IEEEmembership{Senior~Member,~IEEE,} and Tirza Routtenberg,~\IEEEmembership{Senior~Member,~IEEE}}
\thanks{
Parts of this work were presented at the IEEE International Conference on Acoustics, Speech, and Signal Processing (ICASSP) 2023 as the paper~\cite{sagi2023extended}.
 The authors are with the School of ECE, Ben-Gurion University of the Negev, Be'er Sheva, Israel (e-mail: \{itaybuch;sagix;nimrodl;loyay\}@post.bgu.ac.il; \{nirshl; tirzar\}@bgu.ac.il).  
 This work is partially supported by the Israeli Ministry of National Infrastructure, Energy, and Water Resources and by the ISRAEL SCIENCE FOUNDATION (Grant No. 1148/22).
}
\vspace{-0.8cm}
}

\begin{document}
\maketitle
\begin{abstract} 
Dynamic systems of graph signals are encountered in various applications,
including social networks, power grids, and transportation. While such systems can often be described as \ac{ss} models, tracking graph signals via conventional tools based on the \ac{kf} and its variants is typically challenging. This is due to the nonlinearity, high dimensionality, 
irregularity of the domain, and complex modeling associated with real-world dynamic systems of graph signals. In this work, we study the tracking of graph signals using a hybrid model-based/data-driven approach. 
We develop the {\em \namegsp}, which tracks the hidden graphical states from the graphical measurements by jointly leveraging \ac{gsp} tools and \acl{dl} (DL) techniques. 
The derivations of the \namegsp~ are based on extending the KF to exploit the inherent graph structure via graph frequency domain filtering, which considerably simplifies the computational complexity entailed in processing high-dimensional signals and increases the robustness to small topology changes. Then, we use data to learn the \acl{kg} following the recently proposed \acl{kn} framework, which copes with partial and approximated modeling, without forcing a specific model over the noise statistics. Our empirical results demonstrate that the proposed \namegsp~achieves enhanced 
 accuracy and run time performance as well as improved robustness to model misspecifications compared with both model-based and data-
driven benchmarks.
\end{abstract}
 
\acresetall
\vspace{-0.2cm}
\section{Introduction}
\label{sec:intro}
\vspace{-0.1cm}
Graph signals are a fundamental model for describing multivariate signals of irregular and complex structures associated with networks~\cite{8347162,Shuman_Ortega_2013}. These include, for example, transportation systems, social networks, sensor deployments, and power systems. Various applications involve tasks represented as the tracking time sequences of graph signals from nonlinear graph signal observations, representing measurements taken over the network, e.g., in power grids \cite{ramakrishna2021grid,zhao2019learning, giannakis2013monitoring}.

The \ac{kf} \cite{kalman1960new} is the cornerstone algorithm for tracking in dynamic systems. The \ac{kf} relies on the ability to represent the system  as a linear Gaussian \ac{ss} model, while its variants, such as the \ac{ekf}~\cite{larson1967application} and \ac{ukf}~\cite{wan2001unscented}, can track in nonlinear dynamics~\cite{durbin2012time}. However, \ac{kf}-based tracking of graph signals in nonlinear \ac{ss} models is subject to several core limitations. First, the high dimensionality of graph signals significantly increases the computational complexity and latency, and also affects the stability of tracking algorithms. Moreover, \ac{ss} models used in applications involving tracking of graph signals, such as monitoring of power networks~\cite{zhao2016robust, carquex2018state, singh2013decentralized}, are often approximations of complex underlying dynamics. Such mismatches can significantly degrade \ac{kf}-based tracking.

The complexity associated with the high dimensionality of graph signals is typically tackled (though not necessarily in the context of tracking) by exploiting structures in the data arising from the graph/network. A common framework to do so is \ac{gsp}, which extends concepts from traditional digital signal processing for analyzing, representing, and processing data indexed on graphs and signals on networks~\cite{8347162}. GSP leverages the underlying graph structure to capture relationships between data points, while the signals themselves carry meaningful information associated with each node. This framework introduces tools such as the \ac{gft}, graph filters, and methods for the sampling and recovery of graph signals \cite{anis2014towards, marques2015sampling, 8362710,routtenberg2021non}. These tools facilitate the exploitation of the graph structure to propose low-complexity estimators via graph filters that are robust to graph topology changes and mismatches \cite{kroizer2022bayesian, amar2023widely, sagi2022gsp, li2022graph, egilmez2017graph}.

The above estimators consider a single graph signal rather than a time sequence of graph signals. In principle, one can recast a time sequence of graph signals over a fixed time horizon as a   spatio-temporal graph, incorporating both temporal and spatial structures in its graph. As this approach considers the joint processing of a given time frame, it is less suitable for real-time tracking where the goal is to dynamically estimate each incoming observation, as in \ac{kf}-based tracking. 
The \ac{kf} has been employed in several graph-related settings, such as distributed state estimation \cite{Ling_2009, romero2017kernel}, tracking network dynamics \cite{Soule2005, isufi2019forecasting, Isufi2020}, and sampling strategies \cite{di2018adaptive}. However, these techniques are model-based, i.e., they rely on accurate knowledge of the \ac{ss} model, and thus share the sensitivity of the classical \ac{kf} to mismatches and approximation errors.

An alternative tracking approach, which does not rely on \ac{ss} models, learns the scheme from data via machine learning. The growing popularity of \acp{dnn} gave rise to various architectures, such as  \acp{rnn} \cite{gu2017dynamic}, and attention mechanisms \cite{vaswani2017attention}, that are suitable for processing time sequences. Those architectures used to design \ac{dnn}s for tracking, inspired by model-based filtering algorithms \cite{rangapuram2018deep, millidge2021neural, jouaber2021nnakf,becker2019recurrent,klushyn2021latent, jondhale2018kalman}. 
A candidate family of \ac{dnn} architectures for processing graph signals is \acp{gnn}~\cite{jiang2022graph,zhou2020graph, 8579589, parada2021graphon}, which incorporate the graph structure into their learned operation. \acp{gnn} can also be combined with \acp{rnn} for processing time-varying graph signals~\cite{pareja2020evolvegcn,skarding2021foundations}. Nevertheless, data-driven methods, including GNN-based ones, often suffer from difficulties in training, high computational cost, and generalization issues. Moreover, they do not exploite available (though possibly mismatched) knowledge regarding the \ac{ss} model, and operate in a black-box manner, lacking the interpretability of model-based tracking. 

An emerging paradigm for combining model-based signal processing with data-driven approaches is that of model-based \ac{dl}  \cite{he2020model,aggarwal2018modl, shlezinger2022model,shlezinger2023model, nagahama2022graph}.   Model-based \ac{dl} encompasses a set of methodologies for fusing statistical models with \ac{dl} to enhance interpretability, reduce the need for extensive labeled data, and provide reliable predictions \cite{shlezinger2023model}. In the context of tracking in dynamic systems, the  \acl{kn} algorithm proposed in \cite{revach2022kalmannet} augments the \ac{ekf} with a dedicated \ac{rnn}, which enables learning to track in partially known \ac{ss} models while preserving the \ac{ss}-aware \ac{ekf}. 
While its hybrid model-based/data-driven design was shown to improve accuracy under mismatched and nonlinear models~\cite{revach2022kalmannet}, reduce latency~\cite{revach2021rtsnet}, provide confidence measures~\cite{klein2022uncertainty}, and facilitate learning~\cite{revach2022unsupervised}, its complexity significantly increases when processing high-dimensional signals. Existing extensions for mitigating complexity do so by mapping the observations into a low-dimensional subspace~\cite{buchnik2023latent}, thus not handling high-dimensional state settings, as in tracking graph signals in large networks. This motivates accounting for the graph structure in model-based/data-driven tracking of graph signals.

In this paper, we propose {\em \namegsp}, a hybrid data-driven/model-based algorithm for real-time low-complexity tracking of graph signals. \namegsp~is based on the integration of \ac{gsp} properties into the \acl{kn} methodology by identifying the suitable graph filter for handling the complexity and nonlinearity of the \ac{ss} model, while augmenting its operation with a dedicated \ac{rnn} to cope with mismatches and unknown stochasticity. The resulting \namegsp~ is suitable for tracking graph signals in a scalable and low-complexity manner, and learns to operate in possibly mismatched and approximated dynamic modeling.

Specifically, we first assume that the dynamic system of graph signals can be described as a known nonlinear \ac{ss} model, where the main challenge stems from its high dimensionality and nonlinearity. For such settings, we develop GSP-EKF, which alleviates the complexity of the EKF by tracking in the graph frequency domain, while taking the \ac{kg} to be a graph filter. We analyze the GSP-EKF, showcasing the complexity reduction of its \ac{kg} computation, and characterizing analytical conditions for it to achieve the \ac{ekf} performance. By identifying  \ac{gsp}-\ac{ekf} as the suitable model-based algorithm in the presence of full domain knowledge, we follow the \acl{kn} methodology and augment it into \namegsp, addressing partial knowledge of the \ac{ss} model. By identifying the graph filter \ac{kg} of the \ac{gsp}-\ac{ekf} as its specific computation encapsulating the missing domain knowledge, we augment it with a dedicated \ac{rnn} that preserves the scalable graph filter operation. 

We propose a learning algorithm that treats the overall tracking algorithm as a discriminative machine learning architecture~\cite{shlezinger2022discriminative},  training the internal \ac{rnn} of \namegsp~to produce a graph filter that minimizes the $\ell_2$ regularized \ac{mse} in the overall flow of the \ac{gsp}-\ac{ekf}.  Through extensive numerical experiments, we demonstrate the performance and robustness gains of each of the ingredients of \namegsp  compared to both model-based and data-driven benchmarks for synthetic simulations and for power system state estimation (PSSE).

The rest of the paper is organized as follows. Section \ref{sec:System_Model_and_Preliminaries} details \ac{gsp} background, the problem formulation, and briefly describes preliminaries over \ac{ekf} and \acl{kn} algorithms.  Section \ref{sec:Tracking_in_the_Graph_Frequency_Domain}  derives GSP-EKF for scalable tracking in the graph frequency domain, assuming accurate knowledge of the \ac{ss} model. Section \ref{sec:GSP_KalmanNet} presents  \namegsp, including its architecture and training method. Our numerical evaluation is provided in Section \ref{sec:Empirical_Study}, while Section~\ref{sec:conclusions} concludes the paper.

Throughout the paper, we use boldface lower-case letters for vectors (e.g., $\xvec$), boldface upper-case letters for matrices (e.g., ${\bf X}$), and italic letters for sets (e.g., $\mySet{V}$); the $i$th and $(i,j)$th entries of $\xvec$ and ${\bf X}$ are respectively denoted $[\xvec]_i$ and $[{\bf X}]_{i,j}$. The transpose, $\ell_2$ norm, and gradient  are denoted by $\set{\cdot}^\top$,  $\norm{\cdot}$, and  $\nabla_{(\cdot)}$, respectively, while ${\text{diag}}(\xvec)$ is a diagonal matrix with $\xvec$ as its main diagonal. Finally, $\onevec$ is the all-ones vector, and  $\greal$ and $\gint^{+}$ are the sets of real and non-negative integer numbers, respectively.

\vspace{-0.2cm}
\section{System Model and Preliminaries}\label{sec:System_Model_and_Preliminaries}
\vspace{-0.1cm}
In this section, we introduce the considered model of a dynamic system of graph signals, which sets the ground for our derivations of the tracking algorithms in the following sections. In particular, we present the \ac{gsp} notations and graph frequency domain filtering in Subsection~\ref{ssec:Graph_Signal_Processing}. Then, we describe the
\ac{ss} formulation and the considered tracking of a networked data problem in Subsection~\ref{ssec:Problem_Formulation}. Finally, we review the necessary preliminaries for our derivation of \namegsp, which are the model-based \ac{ekf} and data-aided \acl{kn}  of \cite{revach2022kalmannet}, in Subsections~\ref{ssec:EKF} and \ref{ssec:KalmanNet}, respectively.

\vspace{-0.1cm}
\subsection{\ac{gsp} Background} \label{ssec:Graph_Signal_Processing}
\vspace{-0.1cm}
We consider an undirected, connected, weighted graph ${\mySet{G}}(\mySet{V}, \mySet{E},\Wmat)$ representing, e.g., a power grid or a sensor network topology. Here, $\mySet{V}$ and $\mySet{E}$ are sets of nodes (vertices) and edges, respectively, and $\Wmat \in \mathbb{R}^{N \times N}$ is the non-negative weighted adjacency matrix, where $N \define |\mySet{V}|$ is the number of nodes. If there is an edge   $(i,j)\in\mySet{E}$, the entry $[\Wmat]_{i,j}$ represents the weight of the edge; otherwise, $[\Wmat]_{i,j} = 0$.  In general, graph signals are defined as follows~\cite{8347162}:
\begin{definition}[Graph signal]
\label{def:GraphSignal}
A graph signal $\zvec$ with entries in $\mathbb{R}$ is a function $\zvec: \mySet{V} \to \mathbb{R}^N$ defined on the nodes of $\mySet{G}$.
\end{definition}

A common approach to designing filters for graph signals is based on the \ac{gft}. To formulate this, we
use the Laplacian matrix of the graph, which is defined as $\Lmat \define \text{diag} (\Wmat\cdot\onevec) - \Wmat$. 
The \ac{evd}  of $\Lmat$ is given by
 \begin{equation}
\Lmat = \Vmat\Lambdamat \Vmat^{T}.   \label{SVD_new_eq}
 \end{equation}
 In \eqref{SVD_new_eq},  $\Lambdamat$ is a diagonal matrix consisting of the eigenvalues of 
$\Lmat$, $ \lambda_1 < \lambda_2 \leq \ldots \leq \lambda_N $, the  columns of $\Vmat$  are comprised of  the associated eigenvectors, and $\Vmat^{T}=\Vmat^{-1}$. This \ac{evd} is used to define the \ac{gft} over $\mySet{G}$~\cite{Shuman_Ortega_2013}:
\begin{definition}[GFT]
    \label{def:GFT}
    The \ac{gft} of a graph signal defined over a graph $\mySet{G}$ with Laplacian $\Lmat$ is 
     \begin{equation}
\label{GFT}
\tilde{\zvec} \triangleq \Vmat^{T}\zvec. 
 \end{equation}
\end{definition}
Similarly, the inverse \ac{gft} is given by $ \Vmat\tilde{\zvec}$. Linear and shift-invariant graph filters of various graph shift operators play essential roles in \ac{gsp} with various applications \cite{Sandryhaila_Moura_2013}. A common type of such a graph filter, which arises from the \ac{gft} in Definition~\ref{def:GFT}, is the {\em frequency domain graph filter}, defined as follows \cite{ortega2022introduction}:
\begin{definition}
\label{def:GraphFilt}
  A frequency domain graph filter defined over $\mySet{G}$  with Laplacian $\Lmat$ is a function $g(\cdot)$  applied to the Laplacian, which allows the following EVD:
 \begin{equation} \label{laplacian_graph_filter}
  	g(\Lmat)= \Vmat g(\Lambdamat)\Vmat^\top.
  \end{equation}
 where $g(\Lambdamat)$ is a diagonal matrix. In addition, the diagonal elements of $g(\Lambdamat)$ that are associated with   eigenvalues of $\Lmat$ with a multiplicity
greater than 1 are all equal, i.e., if 
 $\lambda_{k+1}=\lambda_k$, then $g(\lambda_{k+1}) = g(\lambda_{k}) $.
\end{definition}

One of the benefits of frequency domain graph filters stems from their reduced complexity when formulating their operation in the \ac{gft} domain. If $\tilde\yvec$ is the output of a frequency domain graph filter applied to a frequency domain graph signal $\tilde\xvec$, i.e., $\tilde\yvec=g(\Lmat)\tilde\xvec$, then it holds that
\begin{equation}
    \label{eqn:FDfilter}
    [\tilde\yvec]_n= g(\lambda_n)   [\tilde\xvec]_n, \quad \forall n \in {1,\ldots,N}.
\end{equation}
where $g(\lambda_n)$ is the graph frequency response of the filter at graph frequency $\lambda_n$. Consequently, instead of filtering with a complexity that grows quadratically with $N$ (multiplying by an $N\times N$ matrix in the vertex domain), it has a complexity that only grows linearly with $N$ via \eqref{eqn:FDfilter}.

\vspace{-0.3cm}
\subsection{Problem Formulation}\label{ssec:Problem_Formulation}
\vspace{-0.1cm}
We consider a dynamic system in discrete-time $t\in\gint^{+}$, with observations $\yvec_t\in \mathbb{R}^N$ and states $\xvec_t\in \mathbb{R}^N$  that are {\em graph signals} over the graph $\mySet{G}$. 
The observations  $\yvec_t$ represent a time sequence measured at each node, while $\xvec_t$ accommodates the latent state sequences at the nodes at time $t$. Accordingly, the  $n$th entry of the  processes $\xvec_{t}$ and $\yvec_t$
creates a time series evolving on node $n$ of the graph \cite{Isufi2020}. While we consider graph signals with scalar values per node, the formulation readily extends to multivariate values at each node.

\subsubsection{\ac{ss} Model}
\label{subsubsection_model}
The dynamics are characterized by a nonlinear, continuous \ac{ss} model in discrete-time:
\begin{subequations}
\label{eqn:ssmodel}
\beqna\label{transition_equation}
&&\xvec_{t} =\fvec_t(\Lmat,\xvec_{t-1}) + \evec_{t},
\\\label{observation_model}
&&\yvec_{t} = \hvec_t(\Lmat,\xvec_{t}) + \vvec_{t}.
\eeqna
\end{subequations}
In \eqref{eqn:ssmodel}, $\evec_{t}\in\mathbb{R}^{N}$  and $\vvec_{t}\in\mathbb{R}^{N}$ are temporally i.i.d., zero-mean, and mutually independent state-evolution and measurement noises, respectively, at time $t$. 
In addition, $\Qmat$ and $\Rmat$ are the covariance matrices of ${\evec}_t$ and $\Vmat_t$, respectively.
The  nonlinear state evolution $\fvec_t:\mathbb{R}^{N \times N} \times{\mathbb{R}}^N \mapsto {\mathbb{R}}^N$  and measurement function $\hvec_t:\mathbb{R}^{N \times N} \times{\mathbb{R}}^N \mapsto {\mathbb{R}}^N$ generally depend on the graph via the Laplacian matrix, $\Lmat$. 

This model comprises a broad family of statistical tracking problems over graphs. For example, a core problem of power system analysis is the recovery of voltages from power measurements, where both the voltages and the powers can be considered as graph signals \cite{routtenberg2021non,drayer2019detection,dabush2023verifying} and the Laplacian matrix is the susceptance matrix. This tracking problem is one of the settings considered in Section \ref{sec:Empirical_Study}. 

\subsubsection{Filtering Problem}
Our goal is to develop a filtering algorithm for tracking the latent graph signal  $\xvec_{t}$ from the observed graph signals $\{\yvec_\tau\}_{\tau\leq t}$ for each time instance $t$. The performance of a given estimator, denoted $\hat{\xvec}_t$, is measured using the average \ac{mse}. That is, for a time sequence of length $T$, the \ac{mse} is computed as
\begin{equation}
\label{eqn:MSE_def}
 {\rm MSE} = \frac{1}{T}\sum_{t=1}^{T}\mathbb{E}\{\|\hat{\xvec}_t - \xvec_t\|^2\}.   
\end{equation}
 
We focus on tracking with {\em partially known \ac{ss} models}. Specifically, we aim to cope with the following challenges:
\begin{enumerate}[label={C.\arabic*}]
\item \label{itmgsp:complexity} The size of the graph signal, $N$, which is the number of nodes, can be very large, causing high complexity and affecting real-time applicability.
\item \label{itmgsp:graph_topology} The graph topology, which can be captured by the Laplacian matrix $\Lmat$, may be misspecified, both in its connectivity pattern and the weights. 
\item \label{itmgsp:Dist} The distributions of the noise signals, $\evec_t$ and $\vvec_t$, are unknown and may be non-Gaussian.
\item \label{itmgsp:Approx} The  state-evolution function, $\fvec(\cdot)$, and the measurement function, $\hvec(\cdot)$, are available, but may be mismatched. 
\end{enumerate}

Challenge~\ref{itmgsp:complexity} is commonly encountered in graph signals due to the growing size of typical virtual and physical networks.   Challenge~\ref{itmgsp:graph_topology} corresponds to the case where one is unaware of the true weights of some of the edges, possibly even wrongly assuming the presence or absence of some edges in a graph.
Challenge~\ref{itmgsp:Dist} stems from the fact that stochasticity in graphical data is often non-Gaussian, while the presence of mismatches in the model functions in Challenge~\ref{itmgsp:Approx} is common in real-world tracking, where the physical phenomena cannot be fully described by simple mathematical equations. 

To cope with the various unknown characteristics of \eqref{eqn:ssmodel} encapsulated in  Challenges~\ref{itmgsp:graph_topology}-\ref{itmgsp:Approx}, we utilize a labeled data-set. The data-set is comprised of $D$ trajectories, which the $d$th trajectory is a sequence of paired graph signals of observations and states in length $T_d$ . The data-set is given by:
\begin{equation}
\label{eqn:DataSet}
    \mySet{D} \triangleq \left\{\left\{\xvec_t^{(d)}, \yvec_t^{(d)}\right\}_{t=1}^{T_d} \right\}_{d=1}^{D}.
\end{equation} 
where $\xvec_t^{(d)}$ and $\yvec_t^{(d)}$ are the state and observation vectors, respectively, at time $t$ and of trajectory $d$.

Our proposed algorithm for tackling Challenges~\ref{itmgsp:complexity}-\ref{itmgsp:Approx} is detailed in Section~\ref{sec:GSP_KalmanNet}. The design follows a model-based \ac{dl} methodology \cite{shlezinger2022model,shlezinger2020model, shlezinger2023model}, where \ac{dl} tools are used to augment and empower existing model-based algorithms though replacing the parts that are relaying on missing domain knowledge with learnable components. Our method builds upon the properties of \ac{gsp}, specifically in the graph frequency domain, 
which enables to reduce the classical \ac{ekf} complexity when dealing with large signals (Challenge \ref{itmgsp:complexity}).
To describe the proposed algorithm, in the following two subsections, we review preliminaries of tracking in partially known \ac{ss} models in a data-driven fashion, building upon the recently proposed \acl{kn} of \cite{revach2022kalmannet} (presented in Subsection~\ref{ssec:KalmanNet}), which augments the classic \ac{ekf} (presented in  Subsection~\ref{ssec:EKF}).  

\subsection{Preliminaries: \ac{ekf}}
\label{ssec:EKF}
Tracking signals in \ac{ss} models in regular domains (i.e.,  not over irregular graphs) is widely studied  (see, e.g., \cite[Ch. 10]{durbin2012time}). One of the most common algorithms, which is suitable for nonlinear models where the noise signals are Gaussian, the \ac{ss} model is fully known, and the dimensions of the system are relatively small, i.e., in the absence of Challenges \ref{itmgsp:complexity}-\ref{itmgsp:Approx}, is the \ac{ekf} \cite{gruber1967approach}. The algorithm follows the operation of the \ac{kf}, combining prediction based on the previous estimate with an update based on the current observation, while extending it to nonlinear \ac{ss} models. 
In particular, the \ac{ekf} first predicts the next state and observation based on  $\hat{\xvec}_{t-1}$ via
\begin{equation}
\label{eqn:Pred}
    \hat{\xvec}_{t|t-1} = \fvec(\hat{\xvec}_{t-1}); \quad \hat{\yvec}_{t|t-1} = \hvec(\hat{\xvec}_{t|t-1}).
\end{equation}
Then, the initial prediction is updated using a  matrix $\Kgain_{t}$, known as the Kalman gain (KG), which dictates the balance between relying on the state evolution function $\fvec(\cdot)$ through \eqref{eqn:Pred} and the current observation $\yvec_t$. 
In particular, the estimator is computed as
\begin{equation}\label{eqn:EKFUpdate}
 \hat{\xvec}_{t} = \Kgain_{t}\cdot \Delta \yvec_t +  \hat{\xvec}_{t|t-1}. 
\end{equation}
where $\Delta \yvec_t \triangleq \yvec_t -  \hat{\yvec}_{t|t-1}$. 
The \ac{kg}  is calculated via 
\begin{equation}
\label{eqn:kalman_gain_computaion}
    \Kgain_{t} = \hat{\bf{\Sigma}}_{t|t-1}\cdot\hat\Hmat_t^\top\cdot \hat{\Smat}_{t|t-1}^{-1}.
\end{equation}
where ${\hat{\bf{\Sigma}}}_{t|t-1}$ and ${\hat{\Smat}}_{t|t-1}$ are the {covariance matrices of the state prediction $\hat{\xvec}_{t|t-1}$ and observation prediction $\hat{\yvec}_{t|t-1}$, respectively. These matrices are calculated via }
\begin{subequations}
    \label{eqn:covariance_computation}
\begin{equation}
\label{eqn:state_covariance_computaion}
{\hat{\bf{\Sigma}}}_{t|t-1}=\hat\Fmat_{t}\cdot\hat{\bf{\Sigma}}_{t-1}\cdot\hat\Fmat_t^\top+\Qmat.
\end{equation}
\begin{equation}
\label{eqn:obs_covariance_computaion}
{\hat\Smat}_{t|t-1}=\hat\Hmat_t\cdot{\hat{\bf{\Sigma}}}_{t|t-1}\cdot\hat\Hmat_t^\top+\Rmat.
\end{equation}
\end{subequations}
where it is assumed that $\hat\Smat_{t|t-1}$ is  a non-singular matrix. The matrices $\hat\Fmat_{t}$ and $\hat\Hmat_{t}$ are the linearized approximations of $\fvec(\cdot)$ and $\hvec(\cdot)$, respectively, obtained using their Jacobian matrices  evaluated at $\hat\xvec_{t-1}$ and $\hat\xvec_{t|t-1}$ (see \cite[Ch. 10]{durbin2012time}), i.e., 
\begin{align}
\label{eqn:Jacobians}
\hat\Fmat_{t}=\nabla_{\xvec}\fvec(\hat\xvec_{t-1}); \quad 
\hat\Hmat_{t}=\nabla_{\xvec}\hvec(\hat{\xvec}_{t|t-1}).
\end{align} 
Finally, the error covariance is updated as follows:
\begin{equation}
\label{eqn:state_covariance_update}
{\hat{\bf{\Sigma}}}_{t}=(\Imat-\Kgain_{t}\cdot\hat\Hmat_{t}){\hat{\bf{\Sigma}}}_{t|t-1}(\Imat-\Kgain_{t}\cdot\hat\Hmat_{t})^\top+\Kgain_{t}\cdot\Rmat\cdot\Kgain_{t}^\top.
\end{equation}

\vspace{-0.5cm}
\subsection{Preliminaries: KalmanNet}
\label{ssec:KalmanNet}
\begin{figure}
\centering\includegraphics[width=0.9\columnwidth]{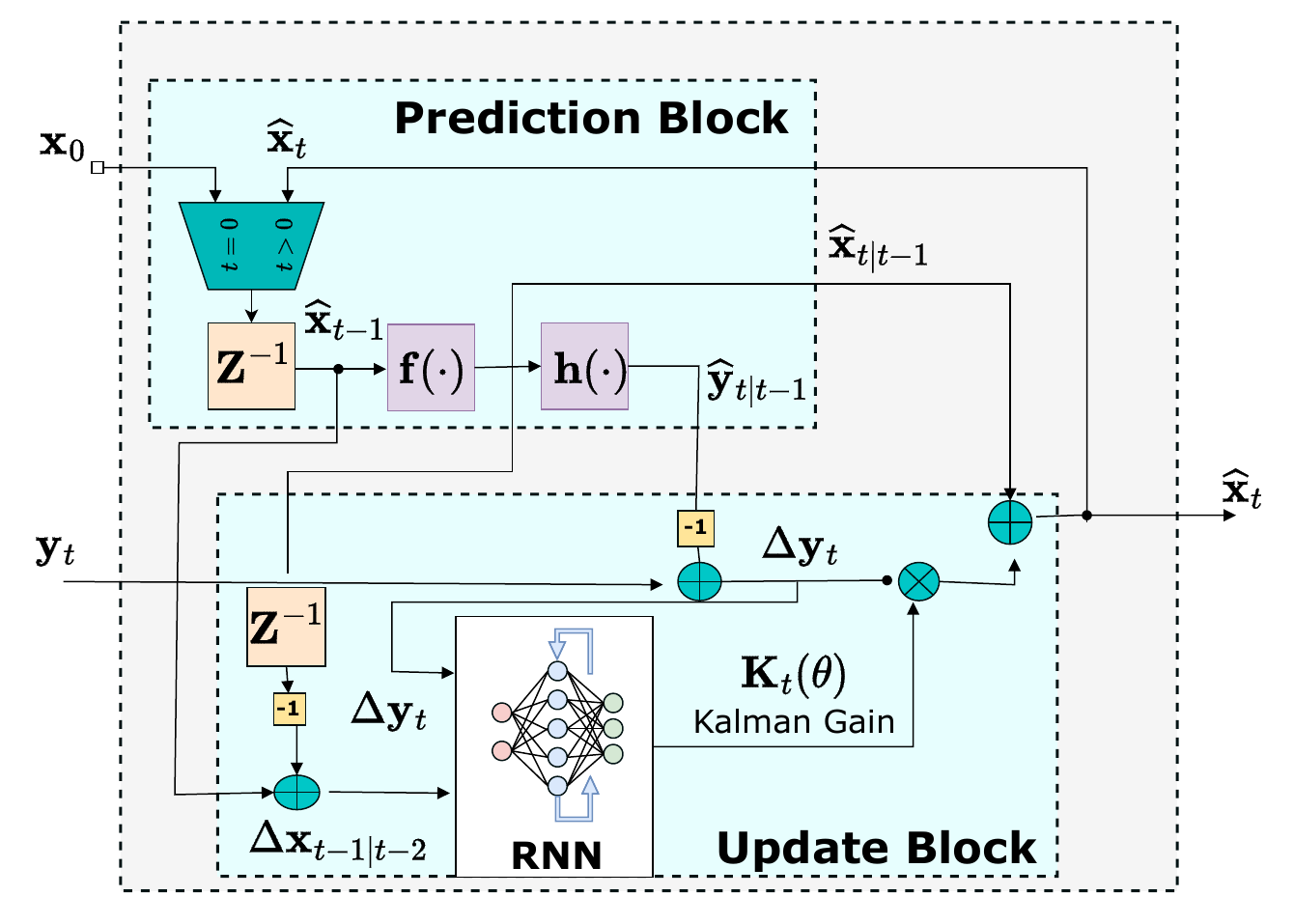}
\caption{KalmanNet block diagram.}\label{fig:kalmanNet_block_diagram}
\end{figure}

Challenges \ref{itmgsp:complexity}-\ref{itmgsp:Approx} significantly limit the applicability of the \ac{ekf} for the setup detailed in Subsection~\ref{ssec:Problem_Formulation}. For instance, the distributions of the noise signals must be known in order to compute the matrices in \eqref{eqn:covariance_computation}.
\acl{kn}, proposed in \cite{revach2022kalmannet}, is a hybrid model-based/data-driven architecture, which leverages data as in \eqref{eqn:DataSet} to tackle Challenges~\ref{itmgsp:Dist}-\ref{itmgsp:Approx} (but not Challenges~\ref{itmgsp:complexity}-\ref{itmgsp:graph_topology}). \acl{kn} builds on the insight that the missing and mismatched domain knowledge is encapsulated in the computation of the \ac{kg}, $\Kgain_{t}$ in \eqref{eqn:kalman_gain_computaion}. Consequently, it augments the \ac{ekf} with a \ac{dl} component by replacing the computation of the \ac{kg} with a dedicated \ac{rnn}-based architecture.
The \ac{kg} is computed using 
the features that are selected as ones that are indicative of the state and measurement noise statistics. For example, these features can be chosen as $\Delta \yvec_t\triangleq \hat{\yvec}_t- \hat{\yvec}_{t|t-1}$ and $\Delta \xvec_t \triangleq \hat{\xvec}_t- \hat{\xvec}_{t-1}$. Accordingly,  \acl{kn} sets the \ac{kg} estimation to be:
\begin{align}
\label{eqn:KGKnet}
\Kgain_t(\KNparams) = \gvec_{\KNparams}^{\rm KG}(\Delta \yvec_t ,\Delta \xvec_{t}), 
\end{align} 
where $\gvec_{\KNparams}^{\rm KG}(\cdot)$ denote the \ac{rnn} mapping with parameters $\KNparams$.
The rest of the filtering operation is the same via \eqref{eqn:Pred}, while replacing the state estimate in \eqref{eqn:EKFUpdate} with
\begin{equation}\label{eqn:KalmaNetUpdate}
 \hat{\xvec}_t = \Kgain_t(\KNparams) \cdot \Delta \yvec_t +  \hat{\xvec}_{t|t-1}.
\end{equation}
The weights of the \ac{rnn}, $\KNparams$, were learned from a data-set \eqref{eqn:DataSet}, based on the regularized \ac{mse} loss function \eqref{eqn:MSE_def}.
During training, the gradients of the loss with respect to $\KNparams$  propagate through the entire \ac{ekf} pipeline to learn the computation of the \ac{kg} as the one yielding the state estimate best matching the data in the sense of the MSE loss. By doing so, \acl{kn} converts the \ac{ekf} into a trainable discriminative model~\cite{shlezinger2022discriminative}, where the data $\mySet{D}$ is used to directly learn the \ac{kg}, bypassing the need to enforce any model over the noise statistics and able to handle domain knowledge mismatches as described in Challenges \ref{itmgsp:Dist}-\ref{itmgsp:Approx}. An illustration of the algorithm block diagram can be seen in Fig.~\ref{fig:kalmanNet_block_diagram}.

\acl{kn} was shown to successfully cope with Challenges~\ref{itmgsp:Dist}-\ref{itmgsp:Approx} when applied to signals with relatively small dimensionality~\cite{revach2022kalmannet}. However, in the considered setup, where the cardinality of the graph signals   (i.e., $N$) can be very large, \acl{kn} may become highly complex (as its number of \ac{rnn} neurons grows with $N^3$~\cite{revach2022kalmannet}), significantly impacting the computational burden and data requirements of its training and inference. Since the state is a high-dimensional graph signal, one cannot mitigate complexity by mapping the observations into a low-dimensional subspace, as done in~\cite{buchnik2023latent, coskun2017long}. This indicates that in order to jointly tackle Challenges~\ref{itmgsp:complexity}-\ref{itmgsp:Approx}, one must account for the graph structure of the signals in the algorithm design, which is the basic principle in our derivation of  \namegsp~in the following sections.

\vspace{-0.25cm}
\section{Tracking in the Graph Frequency Domain}
\label{sec:Tracking_in_the_Graph_Frequency_Domain}
In this section, we design our tracking algorithm for the problem formulated in Subsection~\ref{ssec:Problem_Formulation} in two stages. we first focus only on Challenge~\ref{itmgsp:complexity}, i.e., we assume that the \ac{ss} model is fully known, and thus data is not exploited; then, we extend this approach, by using data and \ac{dl} tools, to also cope with Challenges~\ref{itmgsp:graph_topology}-\ref{itmgsp:Approx} in Section~\ref{sec:GSP_KalmanNet}. We first present the EKF  in the graph frequency domain in Subsection \ref{ssec:EKF_in_the_Graph_Frequency_Domain}, and use this representation to formulate GSP-EKF in Subsection~\ref{ssec:GSP_EKF}, which reduces the computational complexity.
An analysis of the complexity of  GSP-EKF and conditions for coincidence with the EKF are given in Subsection \ref{ssec:Analysis}.

\vspace{-0.2cm}
\subsection{EKF in the Graph Frequency Domain}
\label{ssec:EKF_in_the_Graph_Frequency_Domain}
The EKF for the \ac{ss} model in \eqref{eqn:ssmodel} (assuming known noise distributions) can be implemented in the graph frequency domain. To this end, we transform  \eqref{eqn:ssmodel} to the graph frequency domain by left-multiplying these equations by $\Vmat^{T}$. 
According to \eqref{GFT}, we write the GFTs of the different graph signals as $\tilde{\xvec}_{t} = \Vmat^{T}{\xvec}_{t}$, $\tilde{\evec}_{t} =\Vmat^{T}{\evec}_{t}$, $\tilde{\yvec}_{t} =\Vmat^{T}{\yvec}_{t}$, and $\tilde{\vvec}_{t} =\Vmat^{T}{\vvec}_{t}$.
With a slight abuse of notation, we denote 
\begin{subequations}
\beqna
 \tilde\fvec_{t}(\Lmat,\Vmat\tilde{\xvec}_{t-1})&\triangleq&\Vmat^{T}\fvec_{t}(\Lmat,{\xvec}_{t-1}), \\ \tilde{\hvec}_t(\Lmat,\Vmat\tilde{\xvec}_{t})&\triangleq&\Vmat^\top{\hvec}_t(\Lmat,{\xvec}_{t}).   
 \eeqna
\end{subequations}
Since $\Vmat^\top\Vmat=\Imat$, we obtain the graph frequency \ac{ss} model
\begin{subequations}
 \label{eqn:FDSSmodel}
\beqna\label{transition_equation_graph}
\tilde{\xvec}_{t} &=& \tilde{\fvec}_{t}(\Lmat,\Vmat\tilde{\xvec}_{t-1}) + \tilde{\evec}_{t},
\\\label{observation_model_graph}
\tilde{\yvec}_{t} &=& \tilde{\hvec}_t(\Lmat,\Vmat\tilde{\xvec}_{t}) + \tilde{\vvec}_{t}.
\eeqna
 \end{subequations}
Using  the model assumptions given after \eqref{eqn:ssmodel}, it can be verified that the state-evolution and measurement noises in the graph frequency domain, 
$\tilde{\evec}_{t}\in\mathbb{R}^{N}$  and $\tilde{\vvec}_{t}\in\mathbb{R}^{N}$, respectively, are mutually independent at time $t$, and each one of them
also has independency between different time samples, since they are obtained from the noises in \eqref{eqn:ssmodel} by a linear transformation.  In addition, they still have zero mean and known covariance matrices, that are now given by $\tilde\Rmat \define\Vmat^{T}\Rmat\Vmat$ and $\tilde\Qmat \define\Vmat^{T}\Qmat\Vmat$.

The EKF in the graph frequency domain can be developed directly from the model in \eqref{eqn:FDSSmodel}, in a similar manner to the EKF in Subsection \ref{ssec:EKF}. Alternatively,
since the transformation to the graph frequency domain is an affine unitary transformation, the EKF  in the graph frequency domain can be  obtained from the  standard (vertex-domain) EKF in Subsection~\ref{ssec:EKF} by left-multiplying \eqref{eqn:Pred}-\eqref{eqn:state_covariance_update} by $\Vmat^{T}$, right-multiplying \eqref{eqn:kalman_gain_computaion}-\eqref{eqn:state_covariance_update} by $\Vmat$,
and using Definition~\ref{def:GFT}.
That is,  by   left- and right-multiplying \eqref{eqn:kalman_gain_computaion}  by $\Vmat^{T}$ and $\Vmat$, respectively, we obtain that 
the \ac{kg} of the EKF in the graph frequency domain is given by
		\beqna\label{gain_freq_basic}
    \tilde{\Kgain}_{t} = \tilde{\bf{\Sigma}}_{t|t-1}\cdot\tilde\Hmat_{t}^{T}\cdot\tilde{\Smat}_{t|t-1}^{-1},
\eeqna
where $\tilde{\bf{\Sigma}}_{t|t-1}=\Vmat^{T}\hat{\bf{\Sigma}}_{t|t-1}\Vmat$, $\tilde{\Smat}_{t|t-1}=\Vmat^{T}\hat{\Smat}_{t|t-1}\Vmat$, 
and 
$\tilde\Hmat_{t}=\Vmat^{T}\hat\Hmat_{t}\Vmat$ is the Jacobian matrix within $\hat\Hmat_{t}$ from  \eqref{eqn:Jacobians}, and we used $\Vmat^\top\Vmat=\Imat$. Similarly, \eqref{eqn:state_covariance_update} implies that the error covariance in the graph frequency domain is 
\beqna
 \label{eqn:state_covariance_update2}\tilde{\bf{\Sigma}}_{t|t} = (\Imat-\tilde\Kgain_{t}\cdot\tilde\Hmat_{t})\tilde{\bf{\Sigma}}_{t|t-1}(\Imat-\tilde\Kgain_{t}\cdot\tilde\Hmat_{t})^{T}\hspace{-0.1cm}+\hspace{-0.1cm} \tilde\Kgain_{t}\cdot\tilde\Rmat\cdot\tilde\Kgain_{t}^\top.
 \eeqna
 
The graph frequency domain EKF above is the classic EKF formulated for the \ac{ss} model in \eqref{eqn:FDSSmodel}. Thus, it is still subject to the complexity challenge~\ref{itmgsp:complexity}: computing the \ac{kg} in \eqref{gain_freq_basic} involves inverting an $N\times N$ matrix, and its application involves multiplying the observations and states by an $N\times N$ matrix, $\Vmat^\top$, when transforming to the graph frequency domain. These operations can be computationally costly for large networks, which motivates the derivation of GSP-\ac{ekf}.

\vspace{-0.25cm}
\subsection{GSP-EKF} \label{ssec:GSP_EKF}
The  GSP-EKF algorithm designs the \ac{kg}  as a graph filter.
Thus, we restrict the \ac{kg} to take the form in Definition~\ref{def:GraphFilt}:
\begin{equation} \label{opt_gains}
	{\Kgain_{t}} =  \Vmat g(\Lambdamat )\Vmat^\top.
\end{equation}
where $\Vmat$ and $\Lambdamat$ are obtained from the EVD of
$\Lmat$  in \eqref{SVD_new_eq}, and $g(\cdot)$ is a graph filter, as defined in \eqref{laplacian_graph_filter}.
By \eqref{opt_gains}, this is equivalent to using 
the EKF in the graph frequency domain, while replacing
    $\tilde\Kgain_{t}$ from \eqref{gain_freq_basic} by a diagonal gain:
    \be
    \label{tilde_K}\tilde\Kgain_{t}=g(\Lambdamat ).
    \ee
Substituting \eqref{tilde_K} in \eqref{eqn:state_covariance_update2} leads to the following graph frequency domain  covariance matrix  under graph-filter gain:
\beqna
 \label{update_step_freq2_diag}
 \tilde{\bf{\Sigma}}_{t|t}^{\rm GSP}= (\Imat-g(\Lambdamat )\cdot\tilde\Hmat_{t})\tilde{\bf{\Sigma}}_{t|t-1}(\Imat-g(\Lambdamat )\cdot\tilde\Hmat_{t})^{T} \nonumber\\+ g(\Lambdamat )\cdot\tilde\Rmat\cdot g(\Lambdamat ).\hspace{2.75cm}
\eeqna

Restricting the \ac{kg} to be a graph filter implies that it generally cannot be computed as in the standard EKF, i.e., via \eqref{gain_freq_basic}.
The GSP-EKF thus sets the filter to minimize the computed covariance matrix of the estimator $\hat{\xvec}_t$ over the subset of GSP \acp{kg} in the form of \eqref{tilde_K}. That is,
the  \ac{kg} in the graph frequency domain is set to minimize the trace of \eqref{update_step_freq2_diag}, i.e.,
\beqna
\label{optimal_GSP_kalman_gain}
\tilde\Kgain_{t}^{(\text{GSP})}= \argmin_{g(\Lambdamat )
\in {\DiagSet}_N } \left\{\text{trace}\big( \tilde{\bf{\Sigma}}_{t|t}^{\rm GSP}\big)\right\}, 
\eeqna
where ${\DiagSet}_N$ is the set of diagonal matrices of size $N \times N$.

By substituting \eqref{update_step_freq2_diag} in \eqref{optimal_GSP_kalman_gain}
using the trace operator properties, the minimization from \eqref{optimal_GSP_kalman_gain} can be written  as \beqna\label{trace}
\tilde\Kgain_{t}^{(\text{GSP})}=
    \argmin_{g(\Lambdamat ) \in {\DiagSet}_N}\left\{ \text{trace}((\Imat-g(\Lambdamat )\cdot\tilde\Hmat_{t})\tilde{\bf{\Sigma}}_{t|t-1}\right.\hspace{0.8cm}\nonumber\\\left.
    \cdot(\Imat-g(\Lambdamat )\cdot\tilde\Hmat_{t})^{T})  + \text{trace}(g(\Lambdamat )\cdot\tilde\Rmat\cdot g(\Lambdamat ))\right\}.
\eeqna
By equating the derivative of the r.h.s. of \eqref{trace} with respect to  $g(\Lambdamat )\in {\DiagSet}_N$ to zero, we obtain  that the  solution of \eqref{trace} is
\beqna
\label{opt_sol}
  \tilde\Kgain_{t}^{(\text{GSP})} = \text{ddiag}(\tilde{\bf{\Sigma}}_{t|t-1}\cdot\tilde\Hmat_{t}^\top)\nonumber\hspace{2.75cm}\\\cdot
  (\text{ddiag}(\tilde\Hmat_{t}\cdot\tilde{\bf{\Sigma}}_{t|t-1}\cdot\tilde\Hmat_{t}^{T}+\tilde\Rmat))^{-1},
\eeqna
where  ${\text{ddiag}}(\Dmat)$ is a diagonal matrix with the same diagonal as $\Dmat$.
The GSP-EKF is summarized in Algorithm \ref{alg:GSP-EKF}.
Note that implementing Algorithm \ref{alg:GSP-EKF} while 
replacing $ \tilde\Kgain_{t}^{(\text{GSP})}$ with the \ac{kg} $   \tilde\Kgain_{t} $ in \eqref{gain_freq_basic} results in the EKF in the graph frequency domain from Subsection \ref{ssec:EKF_in_the_Graph_Frequency_Domain}. 

\begin{algorithm}
    \caption{GSP-EKF at time $t$}
    \label{alg:GSP-EKF} 
    \SetKwInOut{Input}{Input}
    \SetKw{KwRet}{Return}
    \Input{
    Laplacian  $\Lmat$; functions $\fvec_t(\cdot)$,  $\hvec_t(\cdot)$;\\
    \ac{ss} model parameters $\Qmat$, and $\Rmat$;}  
        {%
            Compute EVD:  $\Lmat= \Vmat\Lambdamat \Vmat^{T}$;
            
            Compute $\tilde{\Qmat}=\Vmat^{T}\Qmat\Vmat$ and $\tilde\Rmat=\Vmat^{T}{\Rmat}\Vmat$;

Compute linearized $\hat\Fmat_{t}$ and $\hat\Hmat_{t}$ via \eqref{eqn:Jacobians};
            
            Compute $\tilde\Fmat_{t}=\Vmat^{T}\hat\Fmat_{t}\Vmat$ and $\tilde\Hmat_{t}=\Vmat^{T}\hat\Hmat_{t}\Vmat$; 
            
            \underline{Prediction step:}
            \begin{align}
            \label{prediction_step_freq1}
                \hat{\tilde\xvec}_{t|t-1} &=  \Vmat^{T}\fvec_{t}(\Lmat,\Vmat\hat{\tilde\xvec}_{t-1})\\
                \label{prediction_obs_step_freq1}
                \hat{\tilde\yvec}_{t|t-1} &=  \Vmat^{T}\hvec_{t}(\Lmat,\Vmat\hat{\tilde\xvec}_{t|t-1})\\
            \label{prediction_step_freq2}\tilde{\bf{\Sigma}}_{t|t-1} &= \tilde\Fmat_{t}\tilde{\bf{\Sigma}}_{t-1}\tilde\Fmat_{t}^{T} + \tilde{\Qmat}
            \end{align}

            \underline{Update step:}
            
            Compute the \ac{kg} $\tilde\Kgain_{t}^{(\text{GSP})}$ from \eqref{opt_sol}; 
            \begin{align}
            \label{update_step_freq1}
                \hat{\tilde\xvec}_{t} &= \hat{\tilde\xvec}_{t|t-1} + \tilde\Kgain_{t}^{(\text{GSP})}\big(\tilde\yvec_{t}- \hat{\tilde\yvec}_{t|t-1}) \\
            \label{update_step_freq3}
                \tilde{\bf{\Sigma}}_{t} &=(\Imat-\tilde\Kgain_{t}^{(\text{GSP})}\cdot\tilde\Hmat_{t})\tilde{\bf{\Sigma}}_{t|t-1}(\Imat-\tilde\Kgain_{t}^{(\text{GSP})}\cdot\tilde\Hmat_{t})^{T}\nonumber\\&\qquad + \tilde\Kgain_{t}^{(\text{GSP})}\cdot\tilde\Rmat\cdot\tilde\Kgain_{t}^{(\text{GSP})}
            \end{align}       
            }
        \KwRet{$\hat{\xvec}_{t} = \Vmat \hat{\tilde\xvec}_{t}$}
\end{algorithm}

\vspace{-0.5cm}
\subsection{GSP-EKF Discussion}
\label{ssec:Analysis}
\vspace{-0.1cm}
In this subsection, we discuss the properties of the GSP-EKF in Algorithm \ref{alg:GSP-EKF}, including its computational complexity.

\subsubsection{Motivation}
An advantage of implementing the \ac{ekf} in the graph frequency domain, (via the algorithm from Subsection \ref{ssec:EKF_in_the_Graph_Frequency_Domain} or from Subsection \ref{ssec:GSP_EKF}) compared with the \ac{ekf} appears for cases where  $\xvec_t$ and $\evec_{t}$ are
graph bandlimited signals \cite{8347162} with a cutoff graph frequency $N_s$. In these cases,   we can use $[\hat{\tilde{\xvec}}_{t|s}]_n=0$, $\forall n>N_s$, $ t,s \in{\mathbb{Z}}$ and  track only 
the first $N_s$ elements of $\hat{\tilde{\xvec}}_{t}$ at each time step.
Moreover,
the GSP-EKF enforces the \ac{kg} to be a graph filter, which effectively introduces regularization that leverages the underlying graphical information.  This graph filter can be further modified by using parametric graph filter design techniques \cite{sparse_paper}, which enables the development of a distributed, local implementation \cite{5982158}. The integration of graph information via graph filters in the GSP-EKF has the potential to improve the performance in terms of \ac{mse}, interpretability,
robustness, complexity, flexibility, and stability compared to the conventional \ac{ekf}. These advantages have previously been substantiated for static estimation that is based on GSP methodologies \cite{kroizer2022bayesian, sagi2022gsp}.

\subsubsection{Computational complexity} 
The GSP-EKF is particularly designed to cope with the excessive complexity of applying the \ac{ekf} to high-dimensional graph signals (Challenge~\ref{itmgsp:complexity}). We thus commence our analysis of the GSP-EKF with 
 an analysis of the computational savings, as compared with the standard EKF. Unlike the standard EKF, the GSP-EKF  does not require non-diagonal matrix inversion at each time step for computing the \ac{kg}, as in \eqref{eqn:kalman_gain_computaion}. In fact, the \ac{kg} of the GSP-EKF  can be calculated in a vectorized form with a complexity of $\mathcal{O}(N)$ multiplications instead of $\mathcal{O}(N^3)$ (due to inverting an $N\times N$ matrix) for the conventional EKF. While the transformation of the system to the graph frequency domain requires the computation of the EVD of the Laplacian matrix of order $\mathcal{O}(N^3)$,  it can be computed offline, and possibly with low-complexity methods \cite{SVD}. Finally, it should be noted that the transformation of each signal to the graph frequency domain requires multiplication by $N \times N$ matrix, $\Vmat^\top$ $(\mathcal{O}(N^2))$, as opposed to $\mathcal{O}(N^3)$ in the unconstrained EKF.

\subsubsection{Orthogonal graph frequencies}
\label{optimality_conditions}

While the GSP-EKF is in general different from the EKF, in some cases, the filters coincide. One such setting of {\em{orthogonal graph frequencies}} is characterized by the following theorem.
\begin{Theorem} \label{claim_coincides}
The GSP-EKF coincides with the EKF if
\begin{enumerate}
    \item\label{cond1} \hspace{-0.1cm}The matrices, $\Qmat$, and $\Rmat$, are diagonalizable by $\Vmat$. Thus, $\tilde\Qmat$, and $\tilde\Rmat$ are diagonal matrices.
    \item\label{cond2} The functions $\fvec_t(\Lmat,\xvec)$ and $\hvec_t(\Lmat,\xvec)$ are separable in the graph frequency domain, i.e., for all $n=1,\ldots,N$
        \begin{subequations}
            \begin{equation} \label{separately_Model2}
                [\tilde{\fvec}_t(\Lmat,\Vmat\tilde{\xvec})]_n = [\tilde{\fvec}_t(\Lmat,[\tilde{\xvec}]_n\Vmat_{:,n})]_n,
            \end{equation}
            \begin{equation} \label{separately_Model}
                [\tilde{\hvec}_t(\Lmat,\Vmat\tilde{\xvec})]_n = [\tilde{\hvec}_t(\Lmat,[\tilde{\xvec}]_n\Vmat_{:,n})]_n.
            \end{equation}
        \end{subequations}
    where  
    $\Vmat_{:,n}$ is the $n$th column of $\Vmat$, and $[\tilde{\xvec}]_n$ is the $n$th element of $\tilde{\xvec}$
 \end{enumerate}
\end{Theorem}
\begin{IEEEproof}
    The proof is given in the appendix. 
\end{IEEEproof}
\smallskip

Under the conditions of Theorem \ref{claim_coincides}, the \ac{kg} of the conventional EKF  is a graph filter, and the information for estimating the different elements of $\tilde{\xvec}_t$ is decoupled.
Condition \ref{cond1} implies that the elements of the noise signals are uncorrelated in the graph frequency domain for any $t$. Condition \ref{cond2} requires the state evolution and measurement functions to be separable in the graph frequency domain. This situation occurs, e.g., when the states and measurements are linear outputs of graph filters \cite{kroizer2022bayesian}. In addition, the fact that the GSP-EKF coincides with the EKF implies that for the special case of Gaussian linear \ac{ss} models, Algorithm~\ref{alg:GSP-EKF} is reduced to the MSE-optimal Kalman filter, which is equivalent in this case to the graph frequency domain \ac{kf} proposed in \cite{Isufi2020}.

\vspace{-0.2cm}
 \section{\namegsp}
\label{sec:GSP_KalmanNet}
The GSP-EKF derived above only partially copes with the challenges discussed in Subsection~\ref{ssec:Problem_Formulation}. Specifically, it only tackles Challenge \ref{itmgsp:complexity} (Complexity). 
However, its suitability for tracking graph signals indicates that in order to jointly tackle  Challenges~\ref{itmgsp:complexity}-\ref{itmgsp:Approx}, we should leverage data \eqref{eqn:DataSet} and convert the GSP-EKF  into a hybrid discriminative model-based/data-driven architecture~\cite{shlezinger2022discriminative}, in the same spirit of converting the \ac{ekf} to the \acl{kn} in \cite{revach2022kalmannet}).  The resulting \namegsp~algorithm presented in this section is a hybrid, interpretable, and data-efficient scheme for tracking graph signals in partially known nonlinear dynamics.
To introduce \namegsp, we begin by explaining its high level operation in Subsection~\ref{ssec:High_Level_Architecture}, and detail the features processed by its  RNN and its specific architecture in Subsection~\ref{ssec:Neural_Network_Architecture}. Then, we present how \namegsp~is trained in Subsection \ref{ssec:Training_Algorithm}, and provide a discussion in Subsection~\ref{sec:Discussion}.

\vspace{-0.3cm}
\subsection{High Level Architecture}
\label{ssec:High_Level_Architecture}
While the design of \namegsp~follows the methodology used in formulating \acl{kn} (see Subsection~\ref{ssec:KalmanNet}), its architecture differs from that proposed in\cite{revach2022kalmannet}. This stems from the fact that the complexity of \acl{kn}, which is based on the conventional EKF, scales cubically with the dimensions of the signals ($\mathcal{O}(N^3))$, and thus, cannot be used to track graph signals in large networks.  Previous adaptations of \acl{kn} for high-dimensional signals, e.g., \cite{buchnik2023latent}, were tailored for low-dimensional states 
and only high-dimensional observations. 
Accordingly, we derive here the \namegsp~from the GSP-EKF detailed in Subsection~\ref{ssec:GSP_EKF}, leveraging its suitability to track graph signals with affordable complexity, while augmenting its specific computations that are based on unavailable knowledge with trainable components.  

\begin{figure}[hbt]
\includegraphics[width=0.96\columnwidth]{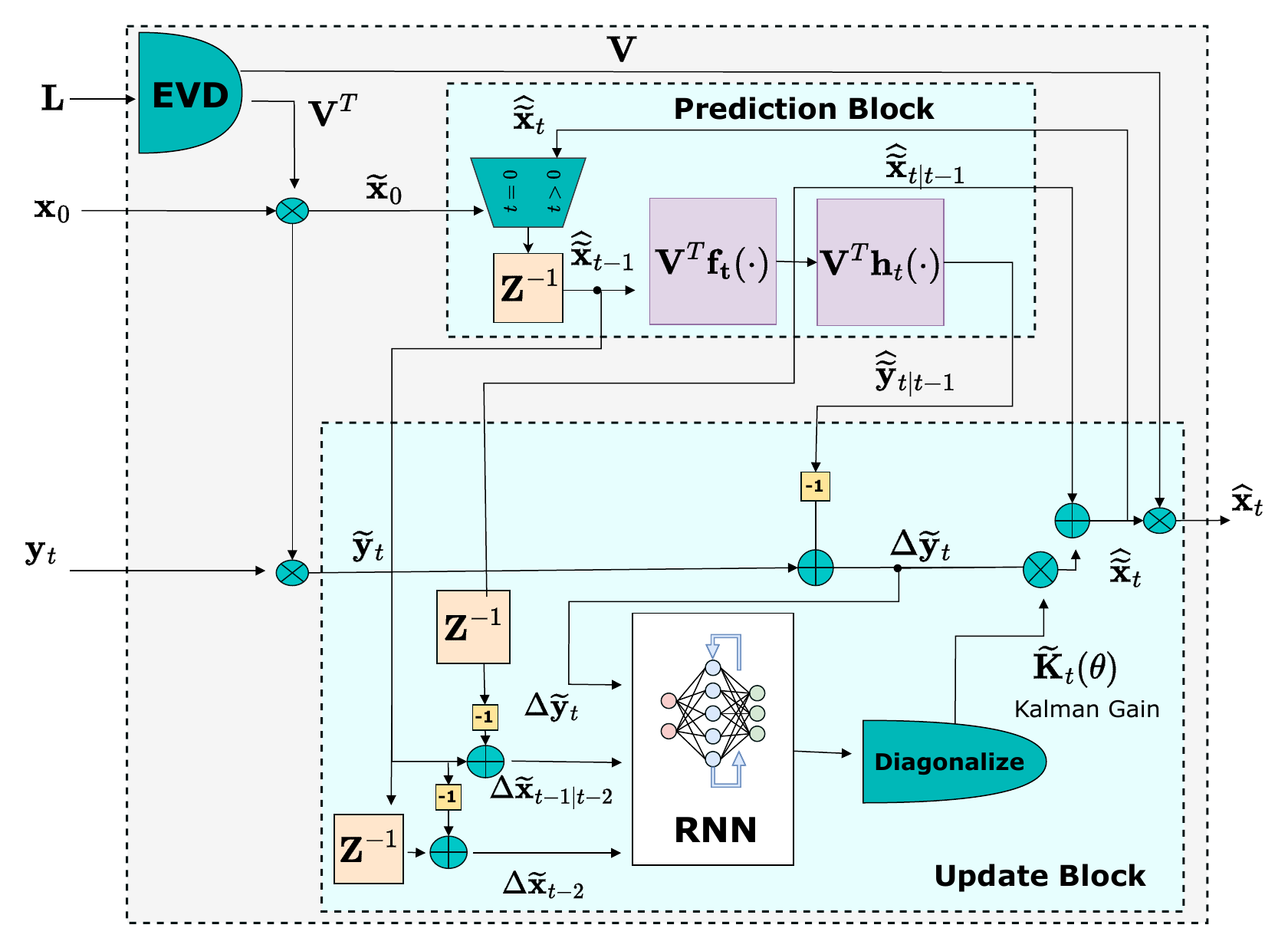}
\caption{\namegsp~block diagram.}
\label{fig:GSP_KalmanNet_block_diagrm}
\end{figure}

As detailed in Subsection~\ref{ssec:EKF_in_the_Graph_Frequency_Domain}, the functions $\tilde\fvec_{t}(\cdot)$ and $\tilde\hvec_{t}(\cdot)$ are known (albeit possibly inaccurately); yet the distribution of the noise terms is unavailable, and potentially non-Gaussian. Their missing statistical moments are used in GSP-EKF only for computing the \ac{kg}. Thus, we design \namegsp~to learn the \ac{kg} from data,
by combining the learned \ac{kg} in the overall GSP-EKF flow. In particular, we use a trainable \ac{rnn} to output the coefficient of a frequency domain graph filter,  enforcing a diagonal matrix manipulation in the graph frequency domain. This architecture is illustrated in Fig.~\ref{fig:GSP_KalmanNet_block_diagrm}. 

For a trained architecture, i.e., when the parameters of the \ac{rnn} denoted $\KNparams$ are set, \namegsp~operates in a similar manner to the GSP-EKF in Algorithm~\ref{alg:GSP-EKF}. Specifically, at each time instance $t$, \namegsp~estimates $\hat{\xvec}_{t}$ in two steps:
\begin{enumerate}
\item In the \emph{prediction} step,
a prior estimator for the current state and observation $\hat{\tilde\xvec}_{t|t-1}$, and $\hat{\tilde\yvec}_{t|t-1}$ are computed from the previous posterior $ \hat{\tilde\xvec}_{t-1}$, in the graph frequency domain by \eqref{prediction_step_freq1G} and \eqref{prediction_step_freq2G} respectively. 
\item In the \emph{update} step, 
the computation of the graph filter comprising the diagonal \ac{kg} is not given explicitly. Rather, it is learned from data using the internal \ac{rnn}, that is thus denoted $\tilde{\Kgain}_t\hspace{-0.04cm}(\KNparams)$, then update the estimate by using the given observation $\tilde{\yvec}_t$ in \eqref{update_step_freq1G}. Unlike GSP-EKF, the inherent memory of \acp{rnn} makes it possible to implicitly track the second-order statistical moments bypassing the need to know the noise distribution.
\end{enumerate}
Since the state is tracked in the graph frequency domain, the vertex domain representation is obtained by the inverse GFT.
This inference procedure of \namegsp~is summarized in Algorithm~\ref{alg:GSP_KalmanNet_Inference}, and is illustrated in the block diagram in Fig.~\ref{fig:GSP_KalmanNet_block_diagrm}.
\vspace{-0.25cm}
\begin{algorithm}
\caption{\namegsp~inference at time $t$}
    \label{alg:GSP_KalmanNet_Inference} 
    \SetKwInOut{Initialization}{Init}
    \SetKw{KwRet}{Return}
    \Initialization{Trained RNN architecture;}
    \SetKwInOut{Input}{Input} 
    \Input{Laplacian  $\Lmat$;
  functions $\fvec_t(\cdot)$, $\hvec_t(\cdot);$}
    {
        {Compute EVD:  $\Lmat= \Vmat\Lambdamat \Vmat^{T}$; \\
        \underline{Prediction step:} 
        \begin{align}
        \label{prediction_step_freq1G}
        \hat{\tilde\xvec}_{t|t-1} &= \Vmat^{T}\fvec_{t}(\Lmat,\Vmat\hat{\tilde\xvec}_{t-1}) \\
        \label{prediction_step_freq2G} \hat{\tilde\yvec}_{t|t-1} &= \Vmat^{T}\hvec_{t}(\Lmat,\Vmat\hat{\tilde\xvec}_{t|t-1}) 
        \vspace{-0.5cm}
        \end{align}
        
        \underline{Update step:} 
        
        Apply the \ac{rnn} to compute \ac{kg} $\tilde{\Kgain}_t(\KNparams)$ by \eqref{eqn:KGKnet};
        \begin{align}
        \label{update_step_freq1G}
        \hat{\tilde\xvec}_{t} = \hat{\tilde\xvec}_{t|t-1} + \tilde\Kgain_{t}(\KNparams)\cdot(\tilde\yvec_{t}-\hat{\tilde\yvec}_{t|t-1})
         \end{align}
        }
    }
    \KwRet{$\hat{\xvec}_{t}=\Vmat\hat{\tilde\xvec}_{t}$}
\end{algorithm}

\vspace{-0.5cm}
\subsection{Neural Network Architecture}
\label{ssec:Neural_Network_Architecture}
The trainable architecture of the internal module at step $4$ of Algorithm \ref{alg:GSP_KalmanNet_Inference}, produces an $N\times 1$ vector comprising the diagonal of the graph frequency domain \ac{kg} denoted $\tilde{\Kgain}_t(\KNparams)$. We use an \ac{rnn} for the training approach, as the recursive nature of the \ac{kg} computation indicates that its learned module should involve an internal memory element. We specifically employ an architecture comprised of a simple \ac{gru} \cite{chung2014empirical} with input shaping and output shaping \ac{fc} layers. The size of the  \ac{gru} hidden state is determined by a factor of $2N$, where $2N$ is the joint dimensionality of the tracked moments $\hat{\tilde{\bf{\Sigma}}}_{t|t-1}$  and $\hat{\tilde{\Smat}}_{t|t-1}$ in \eqref{eqn:state_covariance_computaion} and \eqref{eqn:obs_covariance_computaion}, respectively. 
While the architecture used in Section \ref{sec:Empirical_Study} and presented in Table~\ref{table:RNN_Architecture}  uses 2 \ac{gru} layers, one can utilize multiple layers to increase capacity and abstractness.

Since the computation of the \ac{kg} should be based on features that are indicative of the noise statistics in the \ac{ss}, we use the following signals as the input to the  \ac{rnn}:
\begin{enumerate}[label={\em F\arabic*}]  
\item \label{itmgsp:obDif} \emph{Innovation process} $\Delta\tilde{\yvec}_t={\tilde\yvec_t-\hat{\tilde\yvec}_{t|t-1}}$.
\item \label{itmgsp:FEDif} \emph{Forward evolution difference} $\Delta\hat{\tilde{\xvec}}_{t-2}=\hat{\tilde{\xvec}}_{t-1}-\hat{\tilde{\xvec}}_{t-2}$. This quantity represents the difference between two consecutive posterior state estimates.
\item \label{itmgsp:FUDif} \emph{Forward update difference} $\Delta\tilde\xvec_{t-1|t-2}=\hat{\tilde\xvec}_{t-1}-\hat{\tilde\xvec}_{t-1|t-2}$. This is the difference between the previous posterior state estimator and the prior state estimator.
\end{enumerate}
All these features are in the graph frequency domain, where Feature~\ref{itmgsp:obDif} is informative about the measurement noise, while Features~ \ref{itmgsp:FEDif} and \ref{itmgsp:FUDif} indicate the uncertainty of the state estimator. Accordingly, the overall input is a $3N\times 1$ vector, as presented in Table \ref{table:RNN_Architecture}.
\vspace{-0.15cm}
\begin{table}[hbt]
\centering {
\small
    \begin{tabular}{|c|c|}
    \rowcolor{lightgray}
         \hline
         Layer & Output size \\ [0.7ex]
         \hline
         \hline
         Input & $3N$ \\
         \hline
          FC + ReLU  & $24N$ \\ 
         \hline
          GRU  & $20N$ \\
         \hline
          GRU  & $20N$ \\
         \hline
           FC + ReLU & $4N$ \\ 
         \hline
          FC &  $N$ \\
         \hline
    \end{tabular}}
    \caption{RNN Architecture.}
    \label{table:RNN_Architecture}
\end{table}


\vspace{-0.3cm}
\subsection{Training Algorithm}
\label{ssec:Training_Algorithm}
The trainable component of \namegsp~is the \ac{rnn} outputting the \ac{kg}; the remaining computations follow the GSP-EKF as stated in Algorithm~\ref{alg:GSP_KalmanNet_Inference}. Nonetheless, instead of training the \ac{rnn} based on the gap between its output and a ground truth \ac{kg}, which is unavailable in the presence of Challenges~\ref{itmgsp:graph_topology}-\ref{itmgsp:Approx}, we train the overall algorithm as a discriminative machine learning model. Accordingly, the \ac{rnn} is encouraged to produce a \ac{kg} that is useful in the sense that the resulting state estimator matches the corresponding label in the data-set  \eqref{eqn:DataSet}. We leverage the differentiability of the GSP-EKF operation to enable gradient-based learning via backpropagation through time~\cite{werbos1990backpropagation}.

In particular, we set the loss measure to the  $\ell_2$ regularized \ac{mse}. For a labeled data-set $\mySet{D}$ in \eqref{eqn:DataSet}, the loss is given by 
\begin{align}
\mySet{L}_{\mySet{D}}(\KNparams)=&\frac{1}{|\mySet{D}|}\sum_{d=1}^{|\mySet{D}|} \frac{1}{T_d} \sum_{t=1}^{T_d} \mySet{L}^{(d)}_t(\KNparams) +\lambda \norm{\KNparams}^2,
\label{eqn:loss_function}
\end{align}
with $\lambda>$ 0 being a regularization coefficient. The loss term for each time step $t$ in the $d$th trajectory is obtained via
\begin{equation}
\label{eqn:loss_function_time_step}
\mySet{L}^{(d)}_t(\KNparams) = \norm 
{\hat{\xvec}^{(d)}_t -{{\xvec}}^{(d)}_t(\KNparams)}^2. 
\end{equation}
with $\hat{{\xvec}}^{(d)}_t(\KNparams)$ being the $t$th  estimator of \namegsp~with parameters $\KNparams$ applied to the $d$th  trajectory. 
The resulting procedure optimizing $\KNparams$ based on \eqref{eqn:loss_function} via mini-batch \ac{sgd} is summarized in Algorithm~\ref{alg:training}.

  \begin{algorithm}
    \caption{Training \namegsp}
    \label{alg:training} 
    \SetKwInOut{Initialization}{Init}
    \Initialization{Fix learning rate $\mu>0$; epochs $i_{\max}$; \\  Initial guess $\KNparams$;}
    \SetKwInOut{Input}{Input} 
    \SetKw{KwRet}{Return}
    \Input{
 Laplacian  $\Lmat$; 
 functions $\fvec_t(\cdot)$, $\hvec_t(\cdot)$; \\
    training set  $\mySet{D}$ \eqref{eqn:DataSet};}  
    {

    Compute EVD:  $\Lmat= \Vmat\Lambdamat \Vmat^{T}$; 
    
        \For{$i = 0, 1, \ldots, i_{\max}-1$}
        {%
                    Randomly divide  $\mySet{D}$ into $Q$ batches $\{\mySet{D}_q\}_{q=1}^Q$\;
                    \For{$q = 1, \ldots, Q$}
                    {
                    Apply Algorithm~\ref{alg:GSP_KalmanNet_Inference} to  estimate $\hat{\xvec}_{t}$ \;
                    Compute batch loss $\mathcal{L}_{\mySet{D}_q}(\KNparams)$  by \eqref{eqn:loss_function}\;
                    Update  $\KNparams\leftarrow \KNparams - \mu\nabla_{\KNparams}\mathcal{L}_{\mySet{D}_q}(\KNparams)$\;
                    }
        }
        \KwRet{$\KNparams$}
    }
\end{algorithm}


%

\vspace{-0.1cm}
\subsection{Discussion}
\label{sec:Discussion}
\vspace{-0.1cm}
The proposed 
\namegsp~is particularly tailored for tracking graph signals under the common realistic difficulties highlighted in Challenges~\ref{itmgsp:complexity}-\ref{itmgsp:Approx}. It builds on the understanding that in order to simultaneously cope with incomplete domain knowledge and the inherent complexity associated with high-dimensional graph signals, one cannot rely on conventional EKF and its augmentations, but rather develop a dedicated machine learning architecture that arises from a graph signal-oriented tracking algorithm. For this reason, our first step in deriving \namegsp~is the development of GSP-EKF in Section~\ref{sec:Tracking_in_the_Graph_Frequency_Domain}, followed by its conversion into a trainable algorithm that simultaneously benefits from the available (though possibly approximated) knowledge of the \ac{ss} model alongside learning from data.


Being derived from the GSP-EKF, \namegsp~shares its reduced computational complexity, which is a key component for scalability in light of Challenge~\ref{itmgsp:complexity}. The computational burden of applying a neural network is dictated by its number of weights. 
In \namegsp, computing the Kalman Gain (having $N$ input neurons and $N$ output neurons) has a complexity of $\mathcal{O}(N^2)$, where \acl{kn} involves $\mathcal{O}(N^3)$ computations. Regarding the classical algorithms, EKF is $\mathcal{O}(N^3)$, inverting an $N\times N$ matrix, while GSP-EKF has a complexity of $\mathcal{O}(N^2)$, transforming to the graph frequency domain by multiplying the signals with $N \times N$ matrix. The computational complexity is summarized in Table~\ref{table:benchmarks_comparison}.
Nonetheless, in terms of inference time, as we show in Subsection~\ref{ssec:Latency}, \namegsp~is faster than \acl{kn},  EKF, and GSP-EKF. This stems from the fact that the \ac{dnn}-based methods are amenable to parallelization and acceleration of built-in software accelerators, e.g., PyTorch.

\begin{figure*} 
\begin{center}
\begin{subfigure}[pt]{1.01\columnwidth}
\includegraphics[width=1\columnwidth]{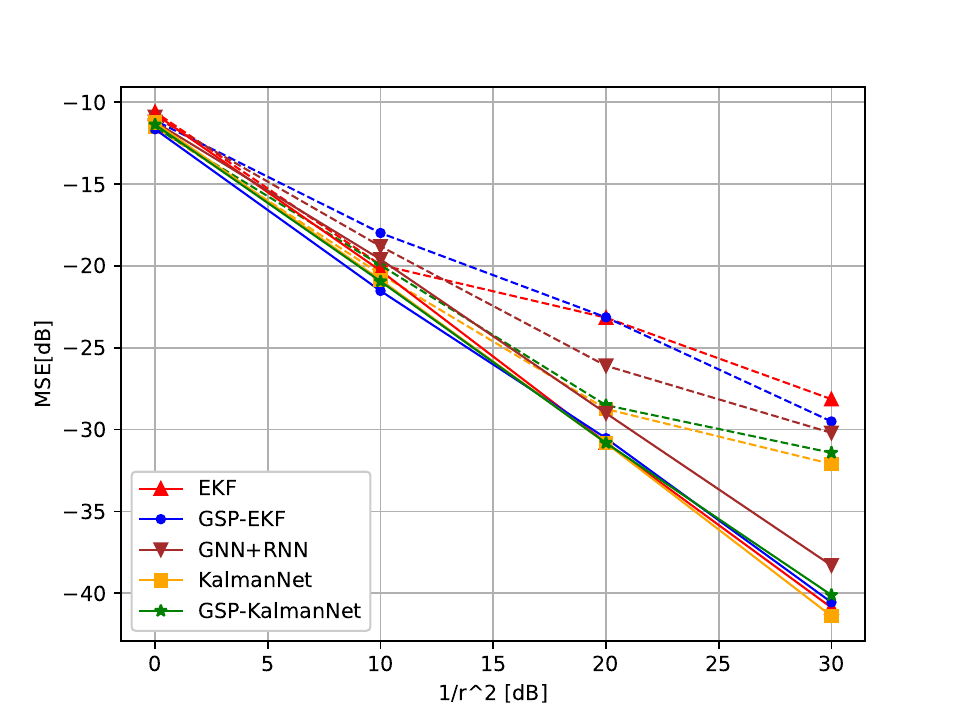}
\caption{Graph with $N=10$ nodes.} 
\label{fig:N=10} 
\end{subfigure}
\begin{subfigure}[pt]{1.01\columnwidth}
\includegraphics[width=1\columnwidth]{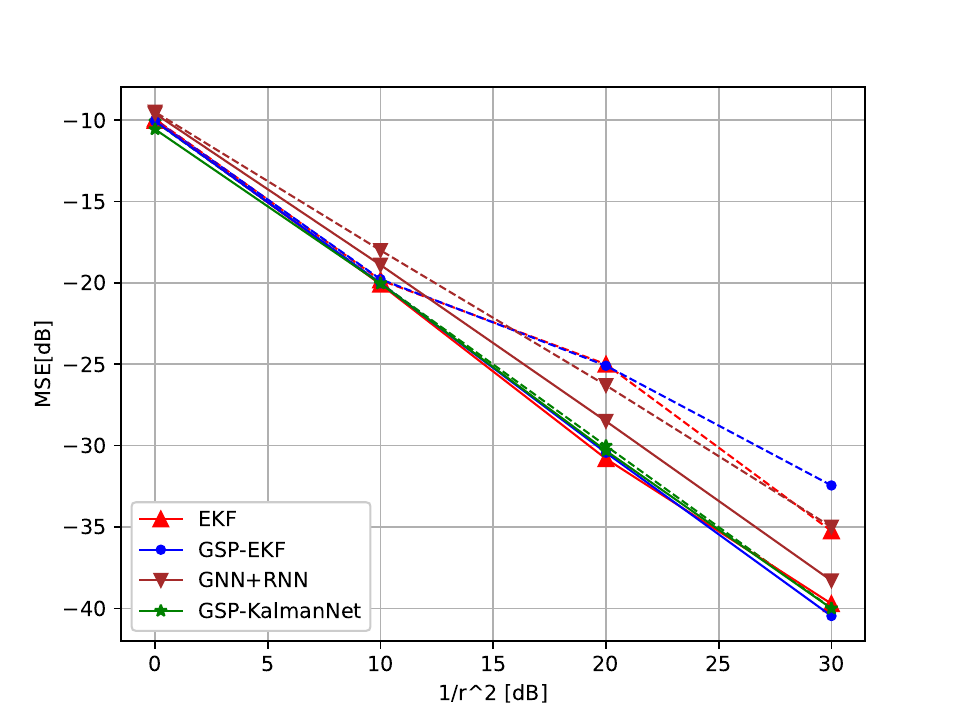}
\caption{Graph with $N=30$ nodes.}
\label{fig:N=30}
\end{subfigure}
\vspace{0.5cm}
\caption{Nonlinear state evolution model: MSE versus inverse observation noise variance,  $\frac{1}{r^2}$.}
\label{fig:Non_linear_state_evolution_function} 
\end{center} 
\vspace{-0.2cm}
\end{figure*}

The ability of \namegsp~to cope with Challenges~\ref{itmgsp:graph_topology}-\ref{itmgsp:Approx} follows directly from its training method and its architecture. Specifically, the fact that training is evaluated based on the estimated state rather than the \ac{kg} (which is the \ac{rnn} output) allows \namegsp~to learn to compute a {\em surrogate} \ac{kg} that enables coping with errors induced by mismatches in $\tilde\fvec_{t}(\cdot)$ and $\tilde\hvec_{t}(\cdot)$, and mispecified graph topology (Challenges~\ref{itmgsp:graph_topology}, \ref{itmgsp:Approx}). In addition, as the \ac{kg} is learned without explicit knowledge of the underlying stochasticity, \namegsp~can learn to track in non-Gaussian noise without knowing its distribution (Challenge~\ref{itmgsp:Dist}). These properties are empirically demonstrated in Section~\ref{sec:Empirical_Study}. Table \ref{table:benchmarks_comparison} details the challenges each benchmark in Section~\ref{sec:Empirical_Study} is facing.

Preserving the model-based operation of the GSP-EKF, as done in \namegsp, can potentially bring operational gains beyond 
increased accuracy and reduced complexity. For instance, it was shown in \cite{klein2022uncertainty} that in some settings one can convert the \ac{kg} into the state error covariance, providing useful uncertainty measures that are typically challenging to acquire in \ac{dl}. Moreover, the fact that the prediction step of the EKF is preserved can be exploited to enable unsupervised learning, as shown in \cite{revach2022unsupervised}. 
Furthermore, the problem formulation in Subsection~\ref{ssec:Problem_Formulation} assumes that the graph is stationary and has a known number of nodes.
A candidate extension of \namegsp~is to deal with graphs with a varying number of nodes and structures, possibly by considering alternative architectures such as \acp{gnn} for computing the \ac{kg}. We leave these investigations and extensions for future work.


\begin{table}
\centering
\caption{Challenges and complexity comparison}
\label{table:benchmarks_comparison}
\centering
\begin{adjustbox}{width=\columnwidth} 
    \begin{tabular}{|c|c|c|c|c|}
    \rowcolor{lightgray}
         \hline
         Challenge & EKF & GSP-EKF & KalmanNet  & \namegsp \\ [0.5ex]
         \hline
         \hline
          \ref{itmgsp:complexity}& X & V & X   & V\\
         \hline
          \ref{itmgsp:graph_topology} & X & V & X   & V \\
            \hline
        \ref{itmgsp:Dist} & X & X & V &   V \\
         \hline
          \ref{itmgsp:Approx} & X & X & V &  V\\
         \hline
           Complexity & $\mathcal{O}(N^3)$ & $\mathcal{O}(N^2)$ & $\mathcal{O}(N^3)$ &   $\mathcal{O}(N^2)$ \\
        \hline
    \end{tabular}
    \end{adjustbox}
    \vspace{-0.5cm}
\end{table}

\vspace{-0.2cm}
\section{Empirical Study}
\label{sec:Empirical_Study}
\vspace{-0.1cm}
In this section, we experimentally evaluate \namegsp\footnote{The source code and hyperparameters used are available at \url{https://github.com/NimrodLeinwand/GSP-KalmanNet}} and compare its performance with those of the filtering model-based/data-driven benchmarks, detailed in the experimental setup in Subsection~\ref{ssec:Experimental_Setup}. We consider cases with both full and partial domain knowledge. In the first study, presented in Subsection~\ref{ssec:Synthetic_data_set}, we consider two synthetic data-sets using nonlinear Gaussian \ac{ss} models. In the second study in Subsection~\ref{ssec:Real_world_scinerio} we consider the practical application domain of power system monitoring~\cite{widely_kalman, zhao2019learning, giannakis2013monitoring} in case of both Gaussian and non-Gaussian noise. We conclude with complexity and latency evaluation in Subsection~\ref{ssec:Latency}.

\vspace{-0.1cm}
\subsection{Experimental Setup}
\label{ssec:Experimental_Setup}
\vspace{-0.1cm}
We simulate dynamic systems that showcase the online tracking capabilities using both synthetic examples and real-world power grid trajectories. In our experiments, we consider both observations and hidden states over a known (though possibly inaccurate) graph structure with $N$ nodes and adjacency matrix ($\Amat$ for an unweighted graph in Subsection \ref{ssec:Synthetic_data_set} and $\Wmat$ for a weighted graph in Subsection \ref{ssec:Real_world_scinerio}). 
The observations are corrupted by two types of measurement noises: 1) An i.i.d. Gaussian noise, $\vvec_t$, with covariance matrix $\Rmat = r^2 \cdot \Imat$; and 
2) An exponential noise with different levels of parameter $\lambda$, used in the setting detailed in Subsection \ref{ssec:Real_world_scinerio}.
The evolution process noise is driven by i.i.d. Gaussian noise $\evec_t$ with covariance matrix $\Qmat = q^2 \cdot \Imat$, setting a fixed ratio of $\log(q^2/r^2) = -20$ dB in the Gaussian case, and $\log(q^2/\lambda) = -20$ dB in the exponential case. 
In each experiment, we generate $D = 2,000$ trajectories of length $T = 200$, forming the data-set $\mathcal{D}$ in \eqref{eqn:DataSet}. Additional $200$ trajectories are reserved for the test set. 

We compare the {\em{GSP-KalmanNet}} with the following  methods:
  $(i)$ \emph{\ac{ekf}} (See Subsection~\ref{ssec:EKF});
  $(ii)$ \emph{GSP-EKF} (See Subsection~\ref{ssec:GSP_EKF});
  $(iii)$  \emph{\acl{kn}} (See Subsection \ref{ssec:KalmanNet});
   $(iv)$ \emph{GNN + RNN}: A \ac{dnn} architecture that combines a \ac{gnn} for spatial information capture with node-wise \ac{rnn} for temporal correlation learning as in~\cite{pareja2020evolvegcn, skarding2021foundations}. This is a baseline data-based method without any assumption on the SS model;
All data-driven methods are trained and tested on the same data-set, with hyperparameters selected to optimize performance.

\vspace{-0.1cm}
\subsection{Synthetic Data-sets}
\label{ssec:Synthetic_data_set}
We first evaluate \namegsp~on two synthetic dynamic systems, utilizing a range of linear and nonlinear functions. For each system, we generate a random unweighted graph using the DGL library~\cite{wang2019deep}, represented by the adjacency matrix $\Amat$.




\subsubsection{Nonlinear State Evolution Model}
\label{sssec:Nonlinear_state_evolution model}
We start our analysis by exploring a straightforward Gaussian \ac{ss} model, with a state-evolution model that takes a nonlinear function with a sinusoidal form influenced by the underlying graph structure: 
\begin{equation}
\label{f_sin}
\fvec_t(\Amat, \xvec_t) = \sin(\xvec_t) + \cos(\xvec_t+\Amat\cdot\xvec_t),
\end{equation}
where $\xvec_t+\Amat\cdot\xvec_t$ is the sum of neighboring nodes up to order 1. The measurement function is the following linear mapping: $\hvec_t(\xvec_t) = 3 \cdot \xvec_t$.
Thus, according to \eqref{eqn:Jacobians}, in this case, $
\hat\Hmat_{t}=3\cdot \Imat$ and $\hat\Fmat_{t}=\frac{\partial \mathbf{f}}{\partial \mathbf{x}_t}$ can be computed from \eqref{f_sin}. 

We analyze two scenarios: \emph{full information}, where the \ac{ss} model parameters (e.g., graph structure) match the one used for data generation, and \emph{partial information}, with mismatched domain knowledge. 
The resulting \acp{mse} 
are presented in Figs.~\ref{fig:N=10}-\ref{fig:N=30} for $N=10$ and $N=30$, respectively, versus $\frac{1}{r^2}$, where $r^2$ is the measurement noise variance.

\begin{figure*} 
\begin{center}
\begin{subfigure}[pt]{1.01\columnwidth}
\includegraphics[width=1\columnwidth]{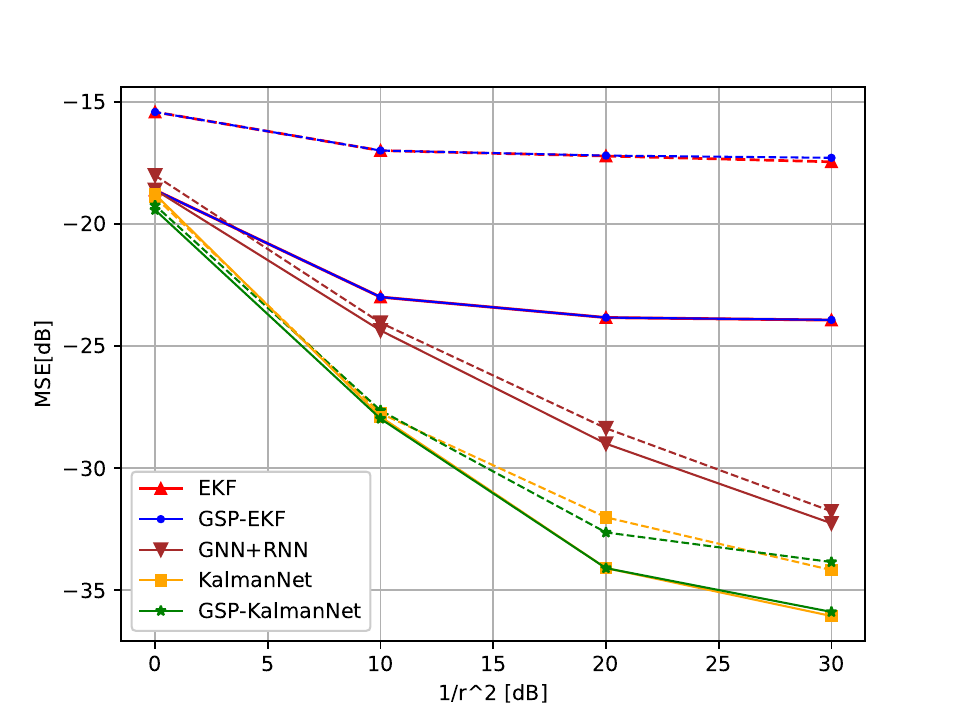}
\caption{MSE versus different Gaussian measurement noise variance $\frac{1}{r^2}$.} 
\label{fig:non_linear_performance} 
\end{subfigure}
\begin{subfigure}[pt]{1.01\columnwidth}
\includegraphics[width=1\columnwidth]{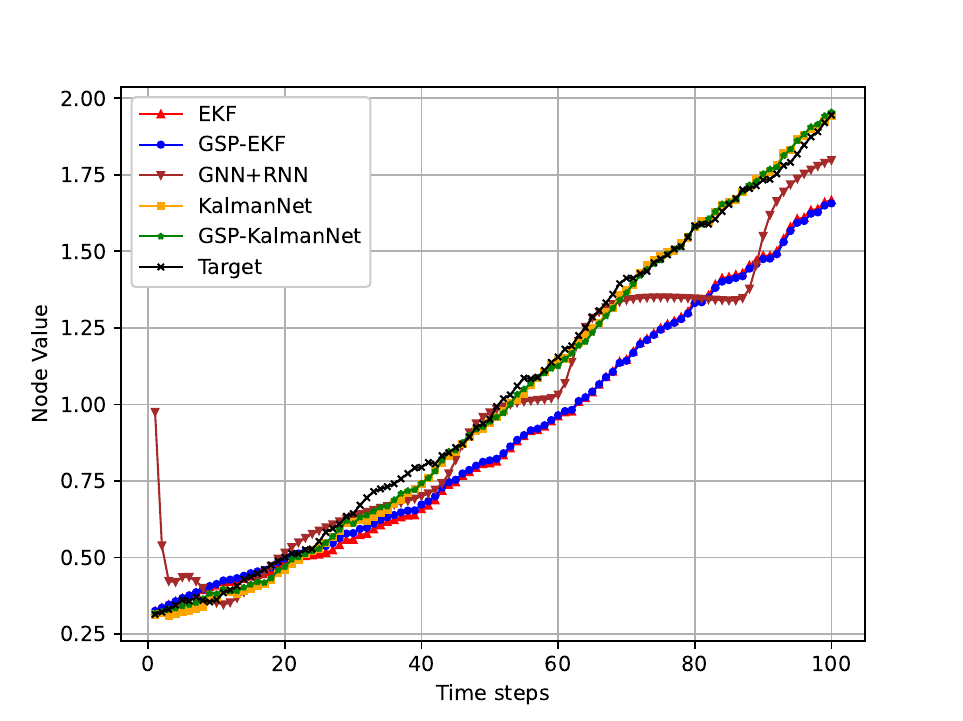}
\caption{State estimation of random node in graph for a single trajectory.}
\label{fig:non_linear_trajectory}
\end{subfigure}
\vspace{0.5cm}
\caption{Nonlinear state evolution and measurement functions, presented in Subsection~\ref{sssec:Non-Linear_model}.}
\label{fig:Nonlinear_model} 
\end{center} 
\vspace{-0.3cm}
\end{figure*}

\textbf{Full Information:} In this setting, all algorithms have access to the dynamic and measurement functions, together with the underlying graph structure (represented by the adjacency matrix $\Amat$) employed during data generation. 
As evident from the solid lines in Fig.~\ref{fig:N=10}, Both \acl{kn},and our proposed \namegsp~successfully learn the model-based mapping, achieving comparable outcomes, while GNN+RNN is slightly behind, not using domain knowledge.

\textbf{Partial Information:} Under partial model information, we evaluate the performance under a mismatch in the adjacency matrix, $\Amat$, having one edge randomly disconnected.  Fig.~\ref{fig:N=10} depicts the performance of the filtering methods in this case as the dashed lines. The trainable algorithms, \acl{kn}, GNN+RNN, and \namegsp, achieve better results compared to the classical methods across a wide range of noise levels. Notably, \namegsp~outperforms \acl{kn} and GNN+RNN in terms of the training process, converges much faster and in a stable manner, which can be attributed to its lower parameter complexity.

Subsequently, we explored the same scenario on a larger unweighted graph with  $N=30$ nodes, each with 10 edges connected to the respective nodes. In this case,  \acl{kn}  exhibits significantly higher RAM requirements for training, rendering it infeasible for training on the same hardware. In Fig.~\ref{fig:N=30} it can be seen that  \namegsp~ outperforms the EKF, GSP-EKF, and data-driven GNN+RNN. Moreover,  \namegsp~successfully learns an adapted \ac{kg} to the setting of incomplete knowledge of the graph structure, achieving the same results as for the case of full domain knowledge. Our simulations (not shown here) present a big gap in the complexity and inference time, as shown in Subsection \ref{ssec:Latency}.

\vspace{0.2cm}
\subsubsection{Nonlinear state evolution and measurement models}
\label{sssec:Non-Linear_model}
Concluding this section, we tackle a  challenging nonlinear Gaussian \ac{ss} model. Here, $\xvec_t$ and $\yvec_t$ represent graph signals of size $N=9$ nodes, where each node is connected to $6$ edges. The state evolution model adopts a sinusoidal form: 
\begin{equation}
\label{f_sin2}
    \fvec_t(\xvec_t) = \xvec_t + \sin(\xvec_t/10 + 3).
\end{equation} 
The measurement model, influenced by the frequency domain, is expressed as 
\begin{equation}
   \hvec_t(\Vmat, \xvec_t) = 0.5{\Vmat\xvec_t} + 0.5 (\Vmat\xvec_t).^3,
\end{equation}
where $(\cdot).^3$ is calculated elementwise. 

We again consider both \emph{full information}, wherein the \ac{ss} model parameters (in this case, both the functions $\fvec(\cdot), \hvec(\cdot)$ and the graph structure that determines $\Vmat$) align with those used for data generation, and \emph{partial information}, wherein the domain knowledge deviates from the ground truth. The resulting \ac{mse} is displayed in Fig.~\ref{fig:non_linear_performance}.

\textbf{Full Information:} Under this setting, we observe in the solid lines in Fig. \ref{fig:non_linear_performance} a gap of more than $10$dB between the \ac{dnn}-aided methods (GNN+RNN, \acl{kn}, and \namegsp) and the model-based ones (EKF and GSP-EKF). Among the data-driven methods, our \namegsp~ achieves the best performance, 
both in terms of \ac{mse}, speed and stability in training, and convergence
from almost any initial weights. Thus, compared to the black-box GNN+RNN, we see how using the state evolution and measurement functions empowers the performance.

\textbf{Partial Information:} 
In order to evaluate the performance of the different methods under partial model information, we assume in this scenario that we are not aware of $2$ random edges in the graph. Furthermore, the state evolution function is characterized by an inaccurate rate (compared with the true model in \eqref{f_sin2}):
\[\fvec_t(\xvec_t) = \xvec_t + \sin(\xvec_t/9 + 3).\] The resulting \acp{mse}, illustrated by the dashed lines in Fig.~\ref{fig:non_linear_performance}, show that the data-driven GNN+RNN achieves a 15dB gap compared to the model-based methods, while the model-based \ac{dl} methods, \acl{kn} and \namegsp, outperform with even more than 20dB. In addition, all three DD methods demonstrate a narrower performance gap compared to their full-information counterparts. Regarding the training process, \namegsp~exhibits again much faster and more stable convergence than \acl{kn} and GNN+RNN, 
which can be attributed to its reduced parameter complexity.
In addition, we illustrate in Fig.~\ref{fig:non_linear_trajectory} the state estimation of a signal at a single node in the graph realization (taking 100 times steps of a smoother state-evolution function for ease of visualization). It can be seen that \namegsp~systematically tracks the ground truth trajectory, compared to its counterparts that deviate from the trajectory. Moreover,   the model-based algorithms, EKF and GSP-EKF, follow the dynamic trend (however, drift slightly) compared to the GNN+RNN that trained only based on data, and curve around the desired trajectory.

\subsection{Power Grid Monitoring - PSSE}
\label{ssec:Real_world_scinerio}
\begin{figure}[hbt]
\centering
\includegraphics[width=0.7\columnwidth]
{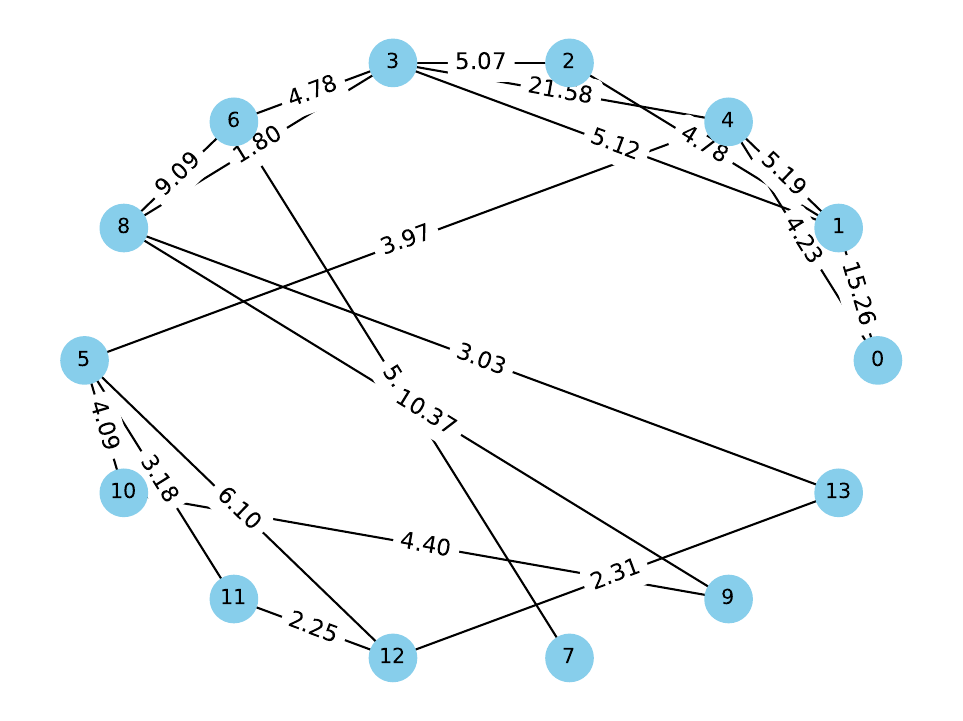}
\caption{Power grid graph with $N=14$ nodes according to the IEEE 14-bus system.}
\label{fig:grapf_structure_power_grid}
\end{figure}

In this subsection, we consider a practical application of power system monitoring \cite{widely_kalman}, with the task of PSSE. A power system can be represented as an undirected weighted graph, ${\mySet{G}}(\mySet{V}, \mySet{E},\Wmat)$, where the set of nodes, $\mySet{V}$, is the set of buses (generators/loads), and the edge set, $\mySet{E}$, is the set of transmission lines between these buses \cite{drayer2019detection, giannakis2013monitoring}. The weight matrix is usually based on the (nodal) admittance matrix of the considered system, which represents the complex weights over the transmission lines.
In particular, in this work, we use the Laplacian matrix that is based on the susceptance matrix, $\Bmat$ (see more in \cite{dabush2023state,halihal2023estimation}).
The graph of $\Lmat$ 
for the IEEE 14-bus system is illustrated in Fig~\ref{fig:grapf_structure_power_grid}. 
\begin{figure*} 
\begin{center}
\begin{subfigure}[pt]{0.99\columnwidth}
\includegraphics[width=1\columnwidth]{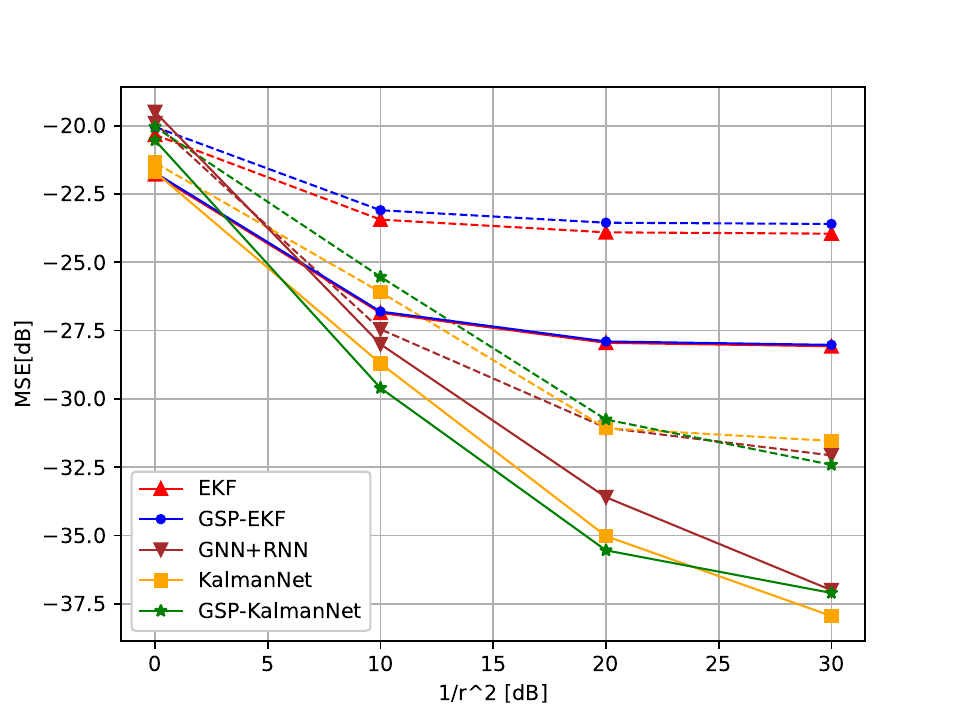}
\caption{Measurement noise $\vvec_t$ is Gaussian} 
\label{fig:power_grid_gaussian_noise} 
\end{subfigure}
\begin{subfigure}[pt]{0.99\columnwidth}
\includegraphics[width=1\columnwidth]{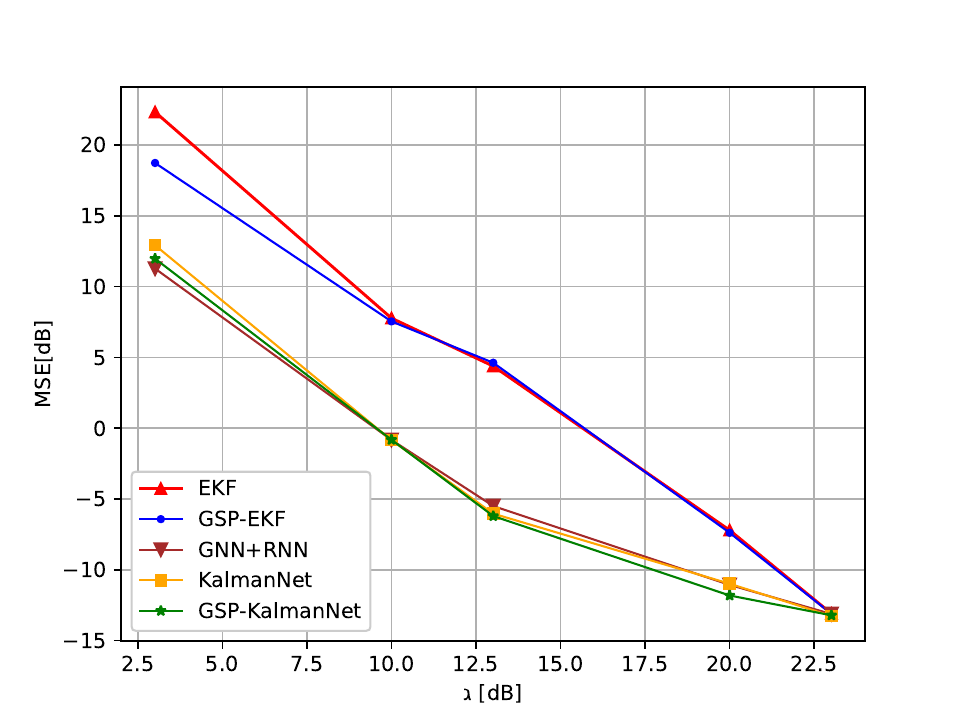}
\caption{Measurement noise $\vvec_t$ is exponential}
\label{fig:power_grid_exponent_noise}
\end{subfigure}
\vspace{0.5cm}
\caption{Power grid data-set performance: MSE versus different Gaussian measurement noise variance $\frac{1}{r^2}$.}
\label{fig:Power_grid_model} 
\end{center} 
\vspace{-0.3cm}
\end{figure*}

In a similar manner to the models in ~\cite{zhao2016robust, carquex2018state, singh2013decentralized},
we use the dynamical model in \eqref{eqn:ssmodel} to describe the state transition and the measurements, where the state vector, $\xvec_t$, consists of the voltage phases of the different buses and $\yvec_t$ contains the active powers at the buses. The commonly-used nonlinear AC power flow model \cite{Abur_book, giannakis2013monitoring}, which is based on Kirchhoff’s and Ohm’s laws, describes the relation between the complex power injection and voltage phasors. According to this model,
the measurement function  can be described by 
\begin{eqnarray} 
\label{h_AC}
\left[\hvec(\xvec_t)\right]_i=\sum\nolimits_{j=1}^N \left([\Gmat]_{i,j}\cos([\xvec_t]_i -[\xvec_t]_j)\right.\hspace{0.25cm}
\nonumber\\\left.+[\Bmat]_{i,j}\sin([\xvec_t]_i -[\xvec_t]_j)\right),
\end{eqnarray}
$i=1,\ldots,N$, 
where $[\Gmat]_{i,j}$ and $[\Bmat]_{i,j}$ are the conductance and susceptance of the transmission line $(i,j)\in\mySet{E}$ \cite{Abur_book}. The goal of dynamic PSSE is to recover the state vector, $\xvec_t$, from the power measurements,
$\yvec_t$, based on \ac{ss} model as in \eqref{eqn:FDSSmodel} with the measurement function from \eqref{h_AC}. In this case, the $(i, j)$th element of  $
\hat\Hmat_{t}=\frac{\partial \mathbf{h}}{\partial \mathbf{x}_t}$ is given by
\[
       [\Gmat]_{i,j} \sin([\xvec_t]_i-[\xvec_t]_j)-[\Bmat]_{i,j}\cos([\xvec_t]_i-[\xvec_t]_j),~~~i \neq j\]
 and 
      \[\sum_{m=1}^N{(-[\Gmat]_{i,m} \sin([\xvec_t]_i-[\xvec_t]_m)+[\Bmat]_{i,m} \cos([\xvec]_i-[\xvec]_m)})\] for $ j = i$.
In particular, since $[\Gmat]_{i,j}=0$  and $[\Bmat]_{i,j}=0$ for any $(i,j) \notin \mySet{E}$, it can be seen that $[\hat{\Hmat}_t]_{i,j}= 0$ for any $(i,j) \notin \mySet{E}$. Thus, $\hat{\Hmat}_t$ is a sparse matrix with the sparsity pattern of the graph. 
In the following, we examine two types of measurement noises: Gaussian and non-Gaussian.


\subsubsection{Gaussian measurement noise}
\label{sssec:Gaussian_observation_noise}
In the first setup, the observations are corrupted by i.i.d. Gaussian noise with different variance value, $r^2$. We use the state-evolution function 
\begin{equation}
\fvec_t(\Wmat, \xvec_t)=1-(\xvec_t + \Wmat\xvec_t).
\end{equation}
influenced by the underlying weighted graph structure with $N=14$ nodes presented in Fig~\ref{fig:grapf_structure_power_grid}, where $\xvec_t + \Wmat\xvec_t$ is the weighted sum of neighboring nodes up to order
1. 
We examine two scenarios: \emph{full information}, wherein \ac{ss} model parameters (the graph structure, depicted in $\Wmat$) align with those used for data generation, and \emph{partial information}, wherein $\Wmat$ is inaccurate. The  \acp{mse} are shown in Fig.~\ref{fig:power_grid_gaussian_noise}.

\textbf{Full Information:} In this context, we observe in the solid lines in Fig. \ref{fig:power_grid_gaussian_noise} a gap of more than 10dB between the learnable methods and the classical ones, while \namegsp~outperforms both \acl{kn} and the purely data-driven GNN+RNN. In this setting, we again
 needed to use numerous initialization weights for \acl{kn} and GNN+RNN, while our \namegsp~ trains in a stable manner for any initialization. 

\textbf{Partial Information:} Assessing \namegsp~under partial model information entails having 10 edges randomly disconnected, which affects the values in the assumed $\Wmat$. The dashed lines in Fig.~\ref{fig:power_grid_gaussian_noise} illustrate that the trainable algorithms achieved an 8dB gap compared to the model-based methods.

\subsubsection{Exponential measurement noise}
\label{sssec:Exponential_observation_noise}
In the second setting, the observations are corrupted by exponential noise with different levels of parameter $\lambda$, indicating the change over time on the noise variable, $\vvec_t$. We use the same measurement function \eqref{h_AC} influenced by the same underlying weighted graph structure from Fig.~\ref{fig:grapf_structure_power_grid}. The state-evolution function in this case is a simple linear mapping
\begin{equation}
\fvec_t(\xvec_t)=\xvec_t+0.05.
\end{equation}

We examine the \emph{full information} case, wherein \ac{ss} model parameters (the graph structure, depicted in $\Gmat$, $\Bmat$ and the functions $\fvec(\cdot)$, $\hvec(\cdot)$) align with those used for data generation. However, the model-based algorithms were given a noise variance for both the state-evolution and measurement noises in the graph
frequency domain $\tilde{\evec}_t$, and $\tilde{\vvec}_t$, assumed to be Gaussian, chosen through a grid search over the data. The \acp{mse} are reported in Fig.~\ref{fig:power_grid_exponent_noise}. We can observe the performance degradation of the model-based methods that have positive dB error for almost all $\lambda$ values. A gap of more than 10dB of performance is achieved by the learnable methods, while \namegsp~outperforms both \acl{kn} and the purely data-driven GNN+RNN. The rightest point shows that when dealing with almost no measurement noise, the data-driven methods learn the same mapping as the classical ones. 

%
%
\vspace{-0.1cm}
\subsection{Complexity and Latency}
\label{ssec:Latency}
We conclude our numerical study by demonstrating that the performance benefits of \namegsp~are achieved with decreased computational complexity and latency compared with model-based \ac{ekf} and the frequency adaption of GSP-EKF, as well as compared to its data-driven counterparts, \acl{kn} and the GNN+RNN architecture. To that end, we provide, in addition to the complexity discussion in Subsection~\ref{sec:Discussion}, an analysis of the average inference time computed over one trajectory. The run time of all tracking algorithms was evaluated on the same platform (Google Colab). For the large graphs, some of the benchmarks exhibited RAM issues.

The data is comprised of  $100$ different trajectories, with $T=200$ time steps each. We used three different graph structures with sizes of $N=10,50, 300$. The resulting average run times are reported in Table~\ref{table:Inference_time_comparison}. As can be seen, the proposed \namegsp, is significantly faster than the benchmark algorithms, especially regarding large graphs.  It stems from the fact that the \ac{ekf} needs to compute the Jacobians and matrix inversions at each time instance to produce its \ac{kg} (Complexity of $\mathcal{O}(N^3)$). Even though GSP-EKF has less complexity (transforming to the graph frequency domain by multiplying with $N \times N$ matrix - ($\mathcal{O}(N^2)$). GSP-EKF is still slower than \namegsp, due to the parallelization and acceleration of built-in software accelerators, e.g., PyTorch. \namegsp~is faster than GNN+RNN since it has fewer weights and it does not crash as \acl{kn}, having a complexity of $\mathcal{O}(N^3)$. These results, combined with the \ac{mse} performance detailed in the previous subsections, demonstrate the ability of \namegsp~in learning to track graph signals while coping with Challenges~\ref{itmgsp:complexity}-\ref{itmgsp:Approx}.


\section{Conclusions}
\label{sec:conclusions}

We proposed a hybrid model-based/data-driven algorithm that tracks graph signals in complex scenarios. Our novel algorithm, \namegsp, leverages the principles of the \ac{ekf} while harnessing the graph structure using \ac{gsp} techniques to simplify the analysis in the graph frequency domain. Furthermore, our method incorporates \ac{dl} components to handle partial and approximate modeling. Our empirical findings demonstrate that \namegsp~outperforms both model-based and data-driven techniques in terms of accuracy, robustness, and run time performance, for tracking high-dimensional graph signals, while coping with the challenges of complexity and partially known dynamics.

\vspace{-0.2cm}
\section*{Appendix A: Proof of Theorem~\ref{claim_coincides}}
Since $\tilde{\Hmat}_{t}=\Vmat^{T}\Hmat_{t}\Vmat$, by using \eqref{eqn:Jacobians}, the GFT definition in \eqref{GFT},  and derivative rules, one obtains
\beqna
\label{H_der}
\tilde\Hmat_{t}=\Vmat^{T}\Hmat_t\Vmat
= \Vmat^{T}\left.\nabla_\xvec \hvec_t(\Lmat,\xvec)\right|_{\xvec=\hat{\xvec}_{t|t-1}}\hspace{-0.15cm}\Vmat
\nonumber\\
= \left.\nabla_{\tilde{\xvec}} \tilde{\hvec}_t(\Lmat,\Vmat\tilde{\xvec})\right|_{\tilde{\xvec}=\hat{\tilde{\xvec}}_{t|t-1}}\Vmat,\hspace{1.85cm}
\eeqna
where $\tilde{\hvec}_t(\Lmat,\Vmat\tilde{\xvec})=\Vmat^\top\hvec_t(\Lmat,\xvec)$.
By substituting \eqref{separately_Model} in \eqref{H_der}, we obtain that under Condition~\ref{cond2}, the r.h.s. of \eqref{H_der} is a diagonal matrix and thus, $\tilde\Hmat_{t}$ from the l.h.s. of \eqref{H_der} is also diagonal. 
Similarly,  \eqref{separately_Model2} implies that $\tilde\Fmat_t$ is also a diagonal matrix.
From Condition \ref{cond1}, $\tilde\Qmat$ is a diagonal matrix. 
Thus, the state covariance estimate of $\tilde{\Sigma}_{t|t-1}$ in \eqref{prediction_step_freq2} is a sum of diagonal matrices for $t=1$. This implies that $\tilde{\bf{\Sigma}}_{1|0}$ is a diagonal matrix. Similarly,  based on Condition \ref{cond1} and \eqref{eqn:obs_covariance_computaion}, 
$\tilde{\Smat}_{1|0}$ is a diagonal matrix. 
Now, one can observe that the \ac{kg} in the graph frequency domain from \eqref{gain_freq_basic}, for $t=1$,  is also diagonal. Therefore, $\tilde\Kgain_{1}=
  \tilde\Kgain_{1}^{(\text{GSP})}$, where $ \tilde\Kgain_{t}^{(\text{GSP})}$ is defined in 
  \eqref{opt_sol}. By substituting these results in \eqref{eqn:state_covariance_update2},
we obtain that $  \tilde{\bf{\Sigma}}_{1|1} $ is   diagonal.
Thus, it can be similarly shown  by induction
 that under Conditions  \ref{cond1}-\ref{cond2},  $  \tilde{\bf{\Sigma}}_{t|t} $ and $\tilde{\bf{\Sigma}}_{t|t-1}$ are diagonal matrices, meaning 
\[ \tilde\Kgain_{t}=
  \tilde\Kgain_{t}^{(\text{GSP})},~~~\forall t\in{\mathbb{Z}}.\]
Thus, the GSP-EKF in Algorithm \ref{alg:GSP-EKF} coincides with the EKF in the graph frequency domain, detailed in Subsection \ref{ssec:EKF_in_the_Graph_Frequency_Domain}, which have the same MSE as the standard EKF presented in Subsection~\ref{ssec:EKF}. Hence, under Conditions \ref{cond1}-\ref{cond2}, the GSP-EKF coincides with the EKF.

\begin{table}
\centering
\small{
\centering
    \begin{tabular}{|c|c|c|c|c|c|}
    \rowcolor{lightgray}
         \hline
          $N$& EKF & GSP-EKF &   \acl{kn} & GNN+RNN & \textbf{Ours} \\ [0.4ex]
         \hline
         \hline
        $10$ & 2.7 & 2.6 & 2.5 & 2.23 & \textbf{1.9} \\
         \hline
         \hline
        $50$ & 3.5 & 3 & 2.7 & 2.4 & \bf{2.1} \\
        \hline
        \hline
        $300$ & -- & 18 & -- & 30 & \bf{3.8} \\
        \hline
    \end{tabular}}
    \caption{Inference run time (in seconds) comparison between the filtering methods over different sizes of graphs. `--' marks infeasible computation due to limited RAM.}\label{table:Inference_time_comparison}
\end{table}

\bibliographystyle{IEEEtran}
\bibliography{IEEEabrv,refs}

\begin{thebibliography}{10}
\providecommand{\url}[1]{#1}
\csname url@samestyle\endcsname
\providecommand{\newblock}{\relax}
\providecommand{\bibinfo}[2]{#2}
\providecommand{\BIBentrySTDinterwordspacing}{\spaceskip=0pt\relax}
\providecommand{\BIBentryALTinterwordstretchfactor}{4}
\providecommand{\BIBentryALTinterwordspacing}{\spaceskip=\fontdimen2\font plus
\BIBentryALTinterwordstretchfactor\fontdimen3\font minus
  \fontdimen4\font\relax}
\providecommand{\BIBforeignlanguage}[2]{{%
\expandafter\ifx\csname l@#1\endcsname\relax
\typeout{** WARNING: IEEEtran.bst: No hyphenation pattern has been}%
\typeout{** loaded for the language `#1'. Using the pattern for}%
\typeout{** the default language instead.}%
\else
\language=\csname l@#1\endcsname
\fi
#2}}
\providecommand{\BIBdecl}{\relax}
\BIBdecl

\bibitem{sagi2023extended}
G.~Sagi, N.~Shlezinger, and T.~Routtenberg, ``Extended {K}alman filter for
  graph signals in nonlinear dynamic systems,'' in \emph{IEEE International
  Conference on Acoustics, Speech and Signal Processing (ICASSP)}, 2023.

\bibitem{8347162}
A.~Ortega, P.~Frossard, J.~Kovačević, J.~M.~F. Moura, and P.~Vandergheynst,
  ``Graph signal processing: Overview, challenges, and applications,''
  \emph{Proc. {IEEE}}, vol. 106, no.~5, pp. 808--828, May 2018.

\bibitem{Shuman_Ortega_2013}
D.~I. Shuman, S.~K. Narang, P.~Frossard, A.~Ortega, and P.~Vandergheynst, ``The
  emerging field of signal processing on graphs: {E}xtending high-dimensional
  data analysis to networks and other irregular domains,'' \emph{{IEEE} Signal
  Process. Mag.}, vol.~30, no.~3, pp. 83--98, May 2013.

\bibitem{ramakrishna2021grid}
R.~Ramakrishna and A.~Scaglione, ``Grid-graph signal processing (grid-{GSP}): A
  graph signal processing framework for the power grid,'' \emph{{IEEE} Trans.
  Signal Process.}, vol.~69, pp. 2725--2739, 2021.

\bibitem{zhao2019learning}
Y.~Zhao, J.~Chen, and H.~V. Poor, ``A learning-to-infer method for real-time
  power grid multi-line outage identification,'' \emph{{IEEE} Trans. Smart
  Grid}, vol.~11, no.~1, pp. 555--564, 2019.

\bibitem{giannakis2013monitoring}
G.~B. Giannakis, V.~Kekatos, N.~Gatsis, S.-J. Kim, H.~Zhu, and B.~F.
  Wollenberg, ``Monitoring and optimization for power grids: A signal
  processing perspective,'' \emph{{IEEE} Signal Process. Mag.}, vol.~30, no.~5,
  pp. 107--128, 2013.

\bibitem{kalman1960new}
R.~E. Kalman, ``A new approach to linear filtering and prediction problems,''
  \emph{Journal of Basic Engineering}, vol.~82, no.~1, pp. 35--45, 1960.

\bibitem{larson1967application}
R.~E. Larson, R.~M. Dressler, and R.~S. Ratner, ``Application of the extended
  {K}alman filter to ballistic trajectory estimation.'' Stanford Research Inst
  Menlo Park CA, Tech. Rep., 1967.

\bibitem{wan2001unscented}
E.~A. Wan and R.~Van Der~Merwe, ``The unscented {K}alman filter,'' \emph{Kalman
  filtering and neural networks}, pp. 221--280, 2001.

\bibitem{durbin2012time}
J.~Durbin and S.~J. Koopman, \emph{Time series analysis by state space
  methods}.\hskip 1em plus 0.5em minus 0.4em\relax OUP Oxford, 2012, vol.~38.

\bibitem{zhao2016robust}
J.~Zhao, M.~Netto, and L.~Mili, ``A robust iterated extended {K}alman filter
  for power system dynamic state estimation,'' \emph{{IEEE} Trans. Power
  Syst.}, vol.~32, no.~4, pp. 3205--3216, 2016.

\bibitem{carquex2018state}
C.~Carquex, C.~Rosenberg, and K.~Bhattacharya, ``State estimation in power
  distribution systems based on ensemble {K}alman filtering,'' \emph{{IEEE}
  Trans. Power Syst.}, vol.~33, no.~6, pp. 6600--6610, 2018.

\bibitem{singh2013decentralized}
A.~K. Singh and B.~C. Pal, ``Decentralized dynamic state estimation in power
  systems using unscented transformation,'' \emph{{IEEE} Trans. Power Syst.},
  vol.~29, no.~2, pp. 794--804, 2013.

\bibitem{anis2014towards}
A.~Anis, A.~Gadde, and A.~Ortega, ``Towards a sampling theorem for signals on
  arbitrary graphs,'' in \emph{IEEE International Conference on Acoustics,
  Speech and Signal Processing (ICASSP)}, 2014.

\bibitem{marques2015sampling}
A.~G. Marques, S.~Segarra, G.~Leus, and A.~Ribeiro, ``Sampling of graph signals
  with successive local aggregations,'' \emph{{IEEE} Trans. Signal Process.},
  vol.~64, no.~7, pp. 1832--1843, 2015.

\bibitem{8362710}
Y.~Tanaka, ``Spectral domain sampling of graph signals,'' \emph{{IEEE} Trans.
  Signal Process.}, vol.~66, no.~14, pp. 3752--3767, 2018.

\bibitem{routtenberg2021non}
T.~Routtenberg, ``Non-{B}ayesian estimation framework for signal recovery on
  graphs,'' \emph{{IEEE} Trans. Signal Process.}, vol.~69, pp. 1169--1184,
  2021.

\bibitem{kroizer2022bayesian}
A.~Kroizer, T.~Routtenberg, and Y.~C. Eldar, ``Bayesian estimation of graph
  signals,'' \emph{{IEEE} Trans. Signal Process.}, vol.~70, pp. 2207--2223,
  2022.

\bibitem{amar2023widely}
A.~Amar and T.~Routtenberg, ``Widely-linear {MMSE} estimation of complex-valued
  graph signals,'' \emph{{IEEE} Trans. Signal Process.}, 2023.

\bibitem{sagi2022gsp}
G.~Sagi and T.~Routtenberg, ``{MAP} estimation of graph signals,'' \emph{arXiv
  preprint arXiv:2209.11638}, 2022.

\bibitem{li2022graph}
P.~Li, N.~Shlezinger, H.~Zhang, B.~Wang, and Y.~C. Eldar, ``Graph signal
  compression by joint quantization and sampling,'' \emph{{IEEE} Trans. Signal
  Process.}, vol.~70, pp. 4512--4527, 2022.

\bibitem{egilmez2017graph}
H.~E. Egilmez, E.~Pavez, and A.~Ortega, ``Graph learning from data under
  {L}aplacian and structural constraints,'' \emph{IEEE Journal of Selected
  Topics in Signal Processing}, vol.~11, no.~6, pp. 825--841, 2017.

\bibitem{Ling_2009}
L.~Shi, ``Kalman filtering over graphs: Theory and applications,'' \emph{IEEE
  Trans. Automatic Control}, vol.~54, no.~9, pp. 2230--2234, 2009.

\bibitem{romero2017kernel}
D.~Romero, V.~N. Ioannidis, and G.~B. Giannakis, ``Kernel-based reconstruction
  of space-time functions on dynamic graphs,'' \emph{{IEEE} J. Sel. Topics
  Signal Process.}, vol.~11, no.~6, pp. 856--869, 2017.

\bibitem{Soule2005}
A.~Soule, K.~Salamatian, A.~Nucci, and N.~Taft, ``Traffic matrix tracking using
  {K}alman filters,'' \emph{Performance Evaluation Review}, vol.~33, pp.
  24--31, 12 2005.

\bibitem{isufi2019forecasting}
E.~Isufi, A.~Loukas, N.~Perraudin, and G.~Leus, ``Forecasting time series with
  {VARMA} recursions on graphs,'' \emph{{IEEE} Trans. Signal Process.},
  vol.~67, no.~18, pp. 4870--4885, 2019.

\bibitem{Isufi2020}
E.~Isufi, P.~Banelli, P.~D. Lorenzo, and G.~Leus, ``Observing and tracking
  bandlimited graph processes from sampled measurements,'' \emph{Signal
  Processing}, vol. 177, 2020.

\bibitem{di2018adaptive}
P.~Di~Lorenzo, P.~Banelli, E.~Isufi, S.~Barbarossa, and G.~Leus, ``Adaptive
  graph signal processing: Algorithms and optimal sampling strategies,''
  \emph{{IEEE} Trans. Signal Process.}, vol.~66, no.~13, pp. 3584--3598, 2018.

\bibitem{gu2017dynamic}
J.~Gu, X.~Yang, S.~De~Mello, and J.~Kautz, ``Dynamic facial analysis: From
  {B}ayesian filtering to recurrent neural network,'' in \emph{IEEE Conference
  on Computer Vision and Pattern Recognition}, 2017, pp. 1548--1557.

\bibitem{vaswani2017attention}
A.~Vaswani, N.~Shazeer, N.~Parmar, J.~Uszkoreit, L.~Jones, A.~N. Gomez,
  {\L}.~Kaiser, and I.~Polosukhin, ``Attention is all you need,''
  \emph{Advances in neural information processing systems}, vol.~30, 2017.

\bibitem{rangapuram2018deep}
S.~S. Rangapuram, M.~W. Seeger, J.~Gasthaus, L.~Stella, Y.~Wang, and
  T.~Januschowski, ``Deep state space models for time series forecasting,''
  \emph{Advances in Neural Information Processing Systems}, vol.~31, 2018.

\bibitem{millidge2021neural}
B.~Millidge, A.~Tschantz, A.~Seth, and C.~Buckley, ``Neural {K}alman
  filtering,'' \emph{arXiv preprint arXiv:2102.10021}, 2021.

\bibitem{jouaber2021nnakf}
S.~Jouaber, S.~Bonnabel, S.~Velasco-Forero, and M.~Pilte, ``{NNAKF}: A neural
  network adapted {K}alman filter for target tracking,'' in \emph{IEEE
  International Conference on Acoustics, Speech and Signal Processing
  (ICASSP)}, 2021, pp. 4075--4079.

\bibitem{becker2019recurrent}
P.~Becker, H.~Pandya, G.~Gebhardt, C.~Zhao, C.~J. Taylor, and G.~Neumann,
  ``Recurrent {K}alman networks: Factorized inference in high-dimensional deep
  feature spaces,'' in \emph{International Conference on Machine
  Learning}.\hskip 1em plus 0.5em minus 0.4em\relax PMLR, 2019, pp. 544--552.

\bibitem{klushyn2021latent}
A.~Klushyn, R.~Kurle, M.~Soelch, B.~Cseke, and P.~van~der Smagt, ``Latent
  matters: Learning deep state-space models,'' \emph{Advances in Neural
  Information Processing Systems}, vol.~34, pp. 10\,234--10\,245, 2021.

\bibitem{jondhale2018kalman}
S.~R. Jondhale and R.~S. Deshpande, ``Kalman filtering framework-based real
  time target tracking in wireless sensor networks using generalized regression
  neural networks,'' \emph{IEEE Sensors Journal}, vol.~19, no.~1, pp. 224--233,
  2018.

\bibitem{jiang2022graph}
W.~Jiang and J.~Luo, ``Graph neural network for traffic forecasting: A
  survey,'' \emph{Expert Systems with Applications}, vol. 207, p. 117921, 2022.

\bibitem{zhou2020graph}
J.~Zhou, G.~Cui, S.~Hu, Z.~Zhang, C.~Yang, Z.~Liu, L.~Wang, C.~Li, and M.~Sun,
  ``Graph neural networks: A review of methods and applications,'' \emph{AI
  open}, vol.~1, pp. 57--81, 2020.

\bibitem{8579589}
F.~Gama, A.~G. Marques, G.~Leus, and A.~Ribeiro, ``Convolutional neural network
  architectures for signals supported on graphs,'' \emph{{IEEE} Trans. Signal
  Process.}, vol.~67, no.~4, pp. 1034--1049, 2019.

\bibitem{parada2021graphon}
A.~Parada-Mayorga, L.~Ruiz, and A.~Ribeiro, ``Graphon pooling in graph neural
  networks,'' in \emph{European Signal Processing Conference (EUSIPCO)}, 2021,
  pp. 860--864.

\bibitem{pareja2020evolvegcn}
A.~Pareja, G.~Domeniconi, J.~Chen, T.~Ma, T.~Suzumura, H.~Kanezashi, T.~Kaler,
  T.~Schardl, and C.~Leiserson, ``Evolve{GCN}: Evolving graph convolutional
  networks for dynamic graphs,'' in \emph{AAAI Conference on Artificial
  Intelligence}, vol.~34, no.~04, 2020, pp. 5363--5370.

\bibitem{skarding2021foundations}
J.~Skarding, B.~Gabrys, and K.~Musial, ``Foundations and modeling of dynamic
  networks using dynamic graph neural networks: A survey,'' \emph{{IEEE}
  Access}, vol.~9, pp. 79\,143--79\,168, 2021.

\bibitem{he2020model}
H.~He, C.-K. Wen, S.~Jin, and G.~Y. Li, ``Model-driven deep learning for mimo
  detection,'' \emph{{IEEE} Trans. Signal Process.}, vol.~68, pp. 1702--1715,
  2020.

\bibitem{aggarwal2018modl}
H.~K. Aggarwal, M.~P. Mani, and M.~Jacob, ``Mo{DL}: Model-based deep learning
  architecture for inverse problems,'' \emph{IEEE Trans. medical imaging},
  vol.~38, no.~2, pp. 394--405, 2018.

\bibitem{shlezinger2022model}
N.~Shlezinger, Y.~C. Eldar, and S.~P. Boyd, ``Model-based deep learning: On the
  intersection of deep learning and optimization,'' \emph{{IEEE} Access},
  vol.~10, pp. 115\,384--115\,398, 2022.

\bibitem{shlezinger2023model}
N.~Shlezinger and Y.~C. Eldar, ``Model-based deep learning,'' \emph{Foundations
  and Trends{\textregistered} in Signal Processing}, vol.~17, no.~4, pp.
  291--416, 2023.

\bibitem{nagahama2022graph}
M.~Nagahama, K.~Yamada, Y.~Tanaka, S.~H. Chan, and Y.~C. Eldar, ``Graph signal
  restoration using nested deep algorithm unrolling,'' \emph{{IEEE} Trans.
  Signal Process.}, vol.~70, pp. 3296--3311, 2022.

\bibitem{revach2022kalmannet}
G.~Revach, N.~Shlezinger, X.~Ni, A.~L. Escoriza, R.~J. Van~Sloun, and Y.~C.
  Eldar, ``Kalman{N}et: Neural network aided {K}alman filtering for partially
  known dynamics,'' \emph{{IEEE} Trans. Signal Process.}, vol.~70, pp.
  1532--1547, 2022.

\bibitem{revach2021rtsnet}
G.~Revach, X.~Ni, N.~Shlezinger, R.~J. van Sloun, and Y.~C. Eldar, ``{RTSN}et:
  Learning to smooth in partially known state-space models,'' \emph{arXiv
  preprint arXiv:2110.04717}, 2021.

\bibitem{klein2022uncertainty}
I.~Klein, G.~Revach, N.~Shlezinger, J.~E. Mehr, R.~J.~G. van Sloun, and Y.~C.
  Eldar, ``Uncertainty in data-driven {K}alman filtering for partially known
  state-space models,'' in \emph{{IEEE} International Conference on Acoustics,
  Speech and Signal Processing {(ICASSP)}}, 2022.

\bibitem{revach2022unsupervised}
G.~Revach, N.~Shlezinger, T.~Locher, X.~Ni, R.~J. van Sloun, and Y.~C. Eldar,
  ``Unsupervised learned {K}alman filtering,'' in \emph{European Signal
  Processing Conference (EUSIPCO)}, 2022, pp. 1571--1575.

\bibitem{buchnik2023latent}
I.~Buchnik, D.~Steger, G.~Revach, R.~J. van Sloun, T.~Routtenberg, and
  N.~Shlezinger, ``Latent-{K}alman{N}et: Learned {K}alman filtering for
  tracking from high-dimensional signals,'' \emph{arXiv preprint
  arXiv:2304.07827}, 2023.

\bibitem{shlezinger2022discriminative}
N.~Shlezinger and T.~Routtenberg, ``Discriminative and generative learning for
  linear estimation of random signals [lecture notes],'' \emph{{IEEE} Signal
  Process. Mag.}, vol.~40, no.~6, pp. 75--82, 2023.

\bibitem{Sandryhaila_Moura_2013}
A.~Sandryhaila and J.~M.~F. Moura, ``Discrete signal processing on graphs,''
  \emph{{IEEE} Trans. Signal Process.}, vol.~61, no.~7, pp. 1644--1656, Apr.
  2013.

\bibitem{ortega2022introduction}
A.~Ortega, ``Introduction to graph signal processing.''\hskip 1em plus 0.5em
  minus 0.4em\relax Cambridge University Press, 2022, ch.~3.

\bibitem{drayer2019detection}
E.~Drayer and T.~Routtenberg, ``Detection of false data injection attacks in
  smart grids based on graph signal processing,'' \emph{{IEEE} Syst. J.},
  vol.~14, no.~2, pp. 1886--1896, 2019.

\bibitem{dabush2023verifying}
L.~Dabush and T.~Routtenberg, ``Verifying the smoothness of graph signals: A
  graph signal processing approach,'' \emph{arXiv preprint arXiv:2305.19618},
  2023.

\bibitem{shlezinger2020model}
N.~Shlezinger, J.~Whang, Y.~C. Eldar, and A.~G. Dimakis, ``Model-based deep
  learning,'' \emph{Proc. {IEEE}}, vol. 111, no.~5, pp. 465--499, 2023.

\bibitem{gruber1967approach}
M.~Gruber, ``An approach to target tracking,'' MIT Lexington Lincoln Lab, Tech.
  Rep., 1967.

\bibitem{coskun2017long}
H.~Coskun, F.~Achilles, R.~DiPietro, N.~Navab, and F.~Tombari, ``Long
  short-term memory {K}alman filters: Recurrent neural estimators for pose
  regularization,'' in \emph{Proceedings of the IEEE International Conference
  on Computer Vision}, 2017, pp. 5524--5532.

\bibitem{sparse_paper}
J.~{Liu}, E.~{Isufi}, and G.~{Leus}, ``Filter design for autoregressive moving
  average graph filters,'' \emph{IEEE Trans. Signal Inf. Process. Netw.},
  vol.~5, no.~1, pp. 47--60, 2019.

\bibitem{5982158}
D.~I. {Shuman}, P.~{Vandergheynst}, and P.~{Frossard}, ``Chebyshev polynomial
  approximation for distributed signal processing,'' in \emph{Proc. of DCOSS},
  2011.

\bibitem{SVD}
L.~Dieci and T.~Eirola, ``On smooth decompositions of matrices,'' \emph{SIAM J.
  on Matrix Anal. and Appl.}, vol.~20, no.~3, pp. 800--819, 1999.

\bibitem{chung2014empirical}
J.~Chung, C.~Gulcehre, K.~Cho, and Y.~Bengio, ``Empirical evaluation of gated
  recurrent neural networks on sequence modeling,'' \emph{arXiv preprint
  arXiv:1412.3555}, 2014.

\bibitem{werbos1990backpropagation}
P.~J. Werbos, ``Backpropagation through time: what it does and how to do it,''
  \emph{Proc. {IEEE}}, vol.~78, no.~10, pp. 1550--1560, 1990.

\bibitem{widely_kalman}
S.~Kanna, D.~H. Dini, Y.~Xia, S.~Y. Hui, and D.~P. Mandic, ``Distributed widely
  linear {K}alman filtering for frequency estimation in power networks,''
  \emph{{IEEE} Trans. Signal Inf. Process. Netw.}, vol.~1, no.~1, pp. 45--57,
  2015.

\bibitem{wang2019deep}
M.~Wang, D.~Zheng, Z.~Ye, Q.~Gan, M.~Li, X.~Song, J.~Zhou, C.~Ma, L.~Yu, and
  Y.~Gai, ``Deep graph library: A graph-centric, highly-performant package for
  graph neural networks,'' \emph{arXiv preprint arXiv:1909.01315}, 2019.

\bibitem{dabush2023state}
L.~Dabush, A.~Kroizer, and T.~Routtenberg, ``State estimation in partially
  observable power systems via graph signal processing tools,'' \emph{Sensors},
  vol.~23, no.~3, p. 1387, 2023.

\bibitem{halihal2023estimation}
M.~Halihal, T.~Routtenberg, and H.~V. Poor, ``Estimation of complex-valued
  laplacian matrices for topology identification in power systems,''
  \emph{arXiv preprint arXiv:2308.03392}, 2023.

\bibitem{Abur_book}
A.~Abur and A.~Gomez-Exposito, \emph{Power System State Estimation: Theory and
  Implementation}.\hskip 1em plus 0.5em minus 0.4em\relax Marcel Dekker, 2004.

\end{thebibliography}

\end{document}